\documentclass[fleqn,usenatbib]{mnras}

\usepackage{newtxtext,newtxmath}

\usepackage[T1]{fontenc}
\usepackage{ae,aecompl}

\usepackage{graphicx}	
\usepackage{amsmath}	
\usepackage{bm}
\newtheorem{theorem}{Definition}

\title[RD Hydro Solver]{A New Residual Distribution Hydrodynamics Solver for Astrophysical Simulations}

\author[B. Morton et al.]{
B. Morton,$^{1}$\thanks{E-mail: morton@roe.ac.uk}
S. Khochfar,$^{1}$
Z. Wu,$^{1}$
\\
$^{1}$Institute for Astronomy, Royal Observatory, Edinburgh EH9 3HJ, UK\\
}

\date{Accepted XXX. Received YYY; in original form ZZZ}

\pubyear{2021}

\begin{document}
\label{firstpage}
\pagerange{\pageref{firstpage}--\pageref{lastpage}}
\maketitle

\begin{abstract}
Many astrophysical systems can only be accurately modelled when the behaviour of their baryonic gas components is well understood. The residual distribution (RD) family of partial differential equation (PDE) solvers produce approximate solutions to the corresponding fluid equations. We present a new implementation of the RD method. The solver efficiently calculates the evolution of the fluid, with up to second order accuracy in both time and space, across an unstructured triangulation, in both 2D and 3D. We implement a novel variable time stepping routine, which applies a drifting mechanism to greatly improve the computational efficiency of the method. We conduct extensive testing of the new implementation, demonstrating its innate ability to resolve complex fluid structures, even at very low resolution. We can resolve complex structures with as few as 3-5 resolution elements, demonstrated by Kelvin-Helmholtz and Sedov blast tests. We also note that we find cold cloud destruction time scales consistent with those predicted by a typical PPE solver, albeit the exact evolution shows small differences. The code includes three residual calculation modes, the LDA, N and blended schemes, tailored for scenarios from smooth flows (LDA), to extreme shocks (N), and both (blended). We compare our RD solver results to state-of-the-art solvers used in other astrophysical codes, demonstrating the competitiveness of the new approach, particularly at low resolution. This is of particular interest in large scale astrophysical simulations, where important structures, such as star forming gas clouds, are often resolved by small numbers of fluid elements.
\end{abstract}




\section{Introduction} \label{sec:intro}

The vast majority of the observed baryonic matter in the Universe is in the form of baryonic gas. The behaviour of this gas can be modelled by solving the Euler equations for an inviscid fluid, which describe the conservation of mass, momentum and energy. These equations must be solved as a set of simultaneous partial differential equations (PDEs). For all but the simplest problems, this must be done numerically. The Navier-Stokes equations, which include the transformation between kinetic and internal energy via viscosity, can also be used. However, the length scales over which this viscosity acts are much smaller than the resolution elements of most astrophysical simulations, allowing the simpler Euler equations to be sufficient, in most cases. Whichever set is used, the equations must be discretised in some manner. Typically this leads to a choice between discretising the problem in space, tracing the fluid evolution using a set of static cells, and discretising the problem by mass \citep{ja:agertz2007}. In the latter case, the gas is modelled as set of massive particles. Astrophysical simulations have been performed with a variety of both these approaches, broadly divided into Eulerian grid based methods \citep{ja:stone1992,ja:stone2008,ja:RAMSES, ja:bryan2014} and Lagrangian particle methods \citep{ja:lucy1977, ja:gingold1977, ja:springel2005}, alongside more recent hybrid moving mesh approaches that combine the Lagrangian nature of the particle methods with the advantages of the Eulerian grids \citep{ja:springel2009, ja:hopkins2015, ja:duffell2016}.  Solving for the evolution of this baryonic gas has been crucial in the development of our understanding of many astrophysical scenarios, from the onset of reionisation and the first galaxies \citep{ja:feng2015, ja:ma2018}, to the evolution of star forming regions \citep{ja:clark2005}, proto-planetary disks \citep{ja:kuffmeier2017}, and so on.

Some of the most successful astrophysical simulation codes solve the evolution of the baryonic gas using Eulerian grids. The majority of these divide the computational domain into a large number of identical cube shaped cells forming a structured mesh. Modern codes often take advantage of mesh refinement algorithms, allowing for cells to be subdivided into multiple smaller cells based on some criterion, typically gas density in astrophysical cases \citep{ja:bryan2014}. This process is known as adaptive mesh refinement (AMR) \citep{ja:berger1984, ja:bryan2014}. The fluid state is traced by these cells, with the evolution found by calculating the flux of gas between finite volume cells. This flux is found by solving the Riemann problem at the cell face \citep{ja:fryxell2000, ja:stone2008}. Riemann solvers innately break the problem down into a set of one dimensional problems across cell faces, which inevitably ignores the information of flows in orthogonal dimensions. This is sometimes referred to as \textit{dimensional splitting}. Flows can only travel across faces, which in structured meshes can lead to preferential flow directions. These can produce numerical artifacts, such as carbuncles, and can suppress flows in other directions. The effect can become more extreme as the resolution increases \citep[see e.g.][]{ja:paardekooper2017}.

Since material cannot flow across the corner of cube cell, corrections can be implemented to attempt to account for these flows \citep{tb:leveque2002}. Over the years, an alternative strategy has emerged, one in which dimensional splitting is not required. These residual distribution (RD) solvers produce truly multi-dimensional hydro-solvers that calculate the evolution of fluids, across some mesh of simplex elements, by calculating the flow across elements in all dimensions at once \citep{ja:abgrall2003, ja:abgrall2006, ja:deconinck2007}. The RD title encompasses a number of different solver formulations.

RD schemes, and their precursors, were developed to solve PDE problems across a host of scales and disciplines. The core advantage of these schemes is their avoidance of dimensional splitting. They solve arbitrary sets of PDEs within domains with potentially complex geometries, and utilising well established, rigorously defined meshes such as the Delaunay triangulation. The current state-of-the-art RD schemes include a number of important characteristics, most notably, allowing for second order accuracy, in both time and space, under certain circumstances \citep{ja:ricchiuto2010, ja:paardekooper2017}. Third and fourth order temporal accuracy has also been achieved \citep{ja:ricchiuto2010}, but the increased computational cost is not justified by the increased accuracy, for the problems we will address here. 

With the advent of these explicit, second order accurate formulations, these methods have emerged as a possible improvement over the traditional fluid methods used in astrophysical simulations. They maintain the strong shock handling capabilities of the Eulerian finite volume  methods, and the related Galerkin finite element schemes \citep{ja:stone2008,ja:skinner2019} (the residual can be formulated analogously to either \citep{ja:deconinck2007,ja:ricchiuto2010}), while combining it with the unstructured and multi-dimensional nature of Lagrangian particle based methods \citep{ja:springel2005,ja:agertz2007}. Astrophysical systems, from large scale structure down to the inter-stellar medium (ISM), contain both highly unstructured flows, and regions with shocks, so combining strengths in resolving both these features is highly desirable, and unlike traditional dimensional splitting Eulerian approaches, there is no additional calculation required when operating on an unstructured mesh. Finally, second order accuracy is achieved on a narrow stencil, requiring only three vertices , when working in two dimensions, or four vertices in three dimensions. The equivalent second order Riemann solver would require four cells in each direction around a face \citep{tb:leveque2002,ja:paardekooper2017}.

We present here a new implementation of an RD solver, built around a static unstructured Delaunay mesh, in both 2D and 3D. The paper is structured as follows. In Section \ref{sec:method} we describe the numerical basis for the RD solver. This includes details of the residual calculation itself, the specifics of the different distribution mechanisms, and our extension to 3D. The implementation is for a static mesh, but the underlying method is naturally suited for future adaptation into such a moving mesh scheme. Section \ref{sec:triangulation} covers the rigorous triangulation method used to build the underlying simplex mesh. In Section \ref{sec:timestep} we detail our novel new variable time-stepping mechanism that allows the residual to be recalculated at different rates in different parts of the mesh, dependent on the local time-step requirement. Section \ref{sec:tests} covers extensive testing of the solver implementation, with tests in 1D, 2D and 3D.

\section{Method} \label{sec:method}

In this section, we lay out the background and derivation of the RD partial differential equation solver \citep{ja:abgrall2006, ja:deconinck2007, ja:ricchiuto2010}. We briefly introduce the precursor to the method, the Roe solver \citep{ja:roe1981, ja:stone2008}. The various choices available as part of the RD approach will also be discussed, including the specific choices made for this implementation.

\subsection{Roe Solver and Residual Distribution in 1D} \label{sec:method_roe}
In order to understand the residual distribution family of methods, it is useful to understand the work that came before their development. The Roe Riemann solver \citep{ja:roe1981, ja:stone2008} is a one dimensional predecessor of the residual distribution family. Roe lays out an approach by which a non-linear system of partial differential equations can be reformulated in a linear form. The method relies on the fact that the solution, to any linear system of hyperbolic partial differential equations (PDEs), can be written as the sum of waves. These waves are the eigenvectors of the Jacobian matrix $\mathbf{A}$ in the set of PDEs, with the wave speeds given by the eigenvalues of this same matrix. This linearisation will be discussed later in Section \ref{sec:method_euler}, but it should be noted that Roe's linearisation for Euler equations, with the appropriate equation of state, is an exact mean-value linearisation, and so will not introduce any additional error \citep{ja:deconinck2007}. For a set of hyperbolic conservation laws of the form
\begin{equation} \label{eq:pde_1D_first}
   \frac{\partial \mathbf{Q}}{\partial t} + \nabla \cdot \mathcal{F}(\mathbf{Q}) = 0,
\end{equation}
where $\mathbf{Q}$ is the state, and $\mathcal{F}(\mathbf{Q})$ is a flux that is a function of the state. In 1D, this is simply the flux $\mathbf{F}(\mathbf{Q})$ in the $x$-direction. It is possible to reformulate this with the Jacobian $\mathbf{A}\equiv\partial \mathbf{F} / \partial \mathbf{Q}$. The original form is a non-linear PDE, but with the substitution, it is now quasi-linear, for a suitably chosen Jacobian $\mathbf{A}$ that is only linearly dependent on the state. The quasi-linear form is
\begin{equation} \label{eq:pde_1D_first_jac}
    \frac{\partial \mathbf{Q}}{\partial t} + \mathbf{A} \frac{\partial \mathbf{Q}}{\partial x} = 0.
\end{equation}
The matrix $\mathbf{A}$ holds all the necessary information to find the solution, as it contains the waves and wave speeds of the fundamental waves of the problem. Roe uses an approximation of the Jacobian from the linearised form at the boundary between two cells \citep{ja:roe1981}, such that the Jacobian represents the flux through the face between the $i-1$ and $i$ cell, giving us
\begin{equation} \label{eq:roe_net_flux}
    \bar{\mathbf{A}}(\mathbf{Q}_{i-1} - \mathbf{Q}_{i}) = \mathbf{F}_{i-1} - \mathbf{F}_{i}.
\end{equation}
The decomposition of the boundary Jacobian reformulates it as a function of it eigenvalues and eigenvectors
\begin{equation}
    \bar{\mathbf{A}} = \mathbf{R}^{-1}\mathbf{\Lambda}\mathbf{R},
\end{equation}
where $\mathbf{R}$ is the matrix constructed by using the eigenvectors of $\bar{\mathbf{A}}$ as columns, and $\Lambda$ is the a diagonal matrix of the corresponding eigenvalues. This decomposition is key to understanding the RD approach, as it breaks down the problem, in all dimensions, into a set of characteristic waves.

The discontinuity at the face can therefore be written as the sum of contributions from its characteristic waves \citep{ja:roe1981}
\begin{equation} \label{eq:roe_disc}
    \mathbf{Q}_{i-1} - \mathbf{Q}_{i} = \sum_{p=1}^{q} \alpha_{i-1/2,p} \mathbf{e}_{i-1/2,p},
\end{equation}
where $\mathbf{\alpha_{i-1/2}}$ is the unknown wave strength of each wave. The eigenvectors are denoted by
$\mathbf{e}_{i-1/2}$, where the $p$ denotes the $p^{th}$ eigenvector. Individual wave strengths are found by projecting the original discontinuity onto the eigenvectors of the Jacobian
\begin{equation}
    \mathbf{\alpha}_{i-1/2} = \mathbf{R}_{i-1/2}^{-1} (\mathbf{Q}_{i-1}-\mathbf{Q}_i),
\end{equation}
where $p$ represents the element of $\mathbf{Q}$, of which there are $q$. It is not necessary to calculate the intermediate states explicitly, since all the information that is required to get the solution to the problem is held in the wave strengths.


The discretised update to the fluid state
\begin{equation}
    \mathbf{Q}_i^{n+1} = \mathbf{Q}_i^n - \frac{\Delta t}{\Delta x} \sum_{p=1}^{q} (\mathbf{\lambda}^p)^+ \alpha^p \mathbf{e}^p,
\end{equation}
where $(\mathbf{\lambda}^p)^+$ is the positive eigenvalue only. Negative eigenvalues are replaced by zero, effectively upwinding the solution. The construction of the Roe method centres around finding the suitable approximation of the Jacobian matrix, in our case, for the inviscid Euler equations. \citep{ja:roe1981}.

The Roe method is essentially the 1D residual distribution method, which can be seen by recasting the net flux at a given boundary as the \textit{residual} material that is left when the flow from one side is combined with material flow from the other. This residual $\phi^T$ is therefore defined as
\begin{equation}
    \phi^T = \mathbf{F}_{i-1} - \mathbf{F}_{i} = \bar{\mathbf{A}}(\mathbf{Q}_{i-1}-\mathbf{Q}_{i}).
\end{equation}
The residual is split between the cells either side of the boundary, in such a way that the distribution still sums to the original total. As with the Roe scheme, the split is achieved by using the positive and negative eigenvalues. Considering the matrix form of the discontinuity, given in Equation (\ref{eq:roe_disc}), one can see how this residual distribution comes about. The element residual is given by
\begin{equation}
    \phi^T = \bar{\mathbf{A}}(\mathbf{Q}_{i-1}-\mathbf{Q}_{i}) = \mathbf{R} \mathbf{\Lambda} \mathbf{R}^{-1}(\mathbf{Q}_{i-1}-\mathbf{Q}_{i}) = \sum_{p=1}^q \lambda^p \alpha^p \mathbf{e}^p.
\end{equation}
As mentioned above, these are distributed by selecting only positive and negative eigenvalues respectively. Thus the residual distributed to cell $i$ is given by
\begin{equation}
    \phi_i = \phi_i^+ = \bar{\mathbf{A}}^+(\mathbf{Q}_{i-1}-\mathbf{Q}_{i}) = \bar{\mathbf{A}}^+ \bar{\mathbf{A}}^{-1} \phi.
\end{equation}
In this form, the solver can easily be extended to more dimensions. To complete the formulation, the final form of the update to the state in cell $i$ is given by the sum of residuals that are distributed to it from its boundaries
\begin{equation} \label{eq:roe_1D_res}
    \mathbf{Q}_i^{n+1} = \mathbf{Q}_i^n - \frac{\Delta t}{\Delta x} \left(\phi_i^+ + \phi_i^-\right),
\end{equation}
or in other words, the update comes from the summation of the residuals sent from the boundaries.

\subsection{Residual Distribution in 2D} \label{sec:method_rd2D}
Since the original formulation of this method, a significant amount of work has gone in to extending it to higher dimensions \citep{ja:deconinck1993, ja:paillere1995, ja:abgrall2003, ja:abgrall2007, ja:ricchiuto2010}. In this section, we define the generalised form of the residual, in dimensions greater than one.

\subsubsection{Notation and Geometry}
\begin{figure}
    \centering
    \includegraphics[width=0.4\textwidth]{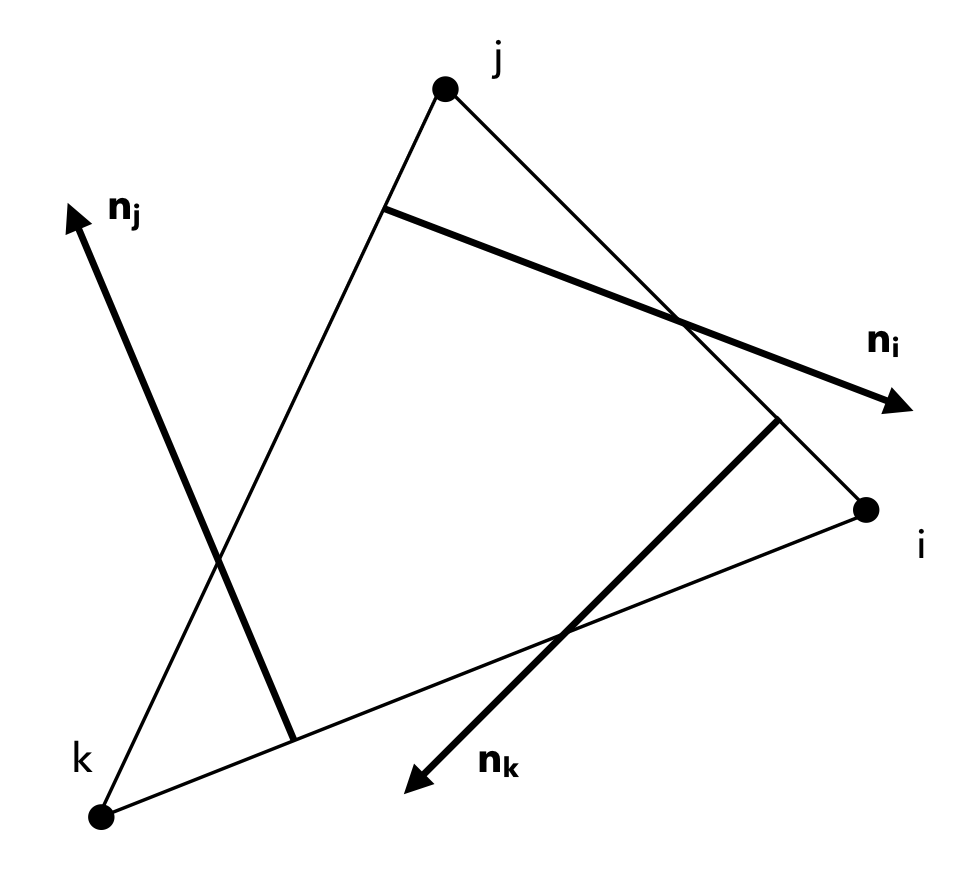}
    \caption{Element vertices and associated normals}
    \label{fig:method_normal}
\end{figure}

\begin{figure}
    \centering
    \includegraphics[width=0.4\textwidth]{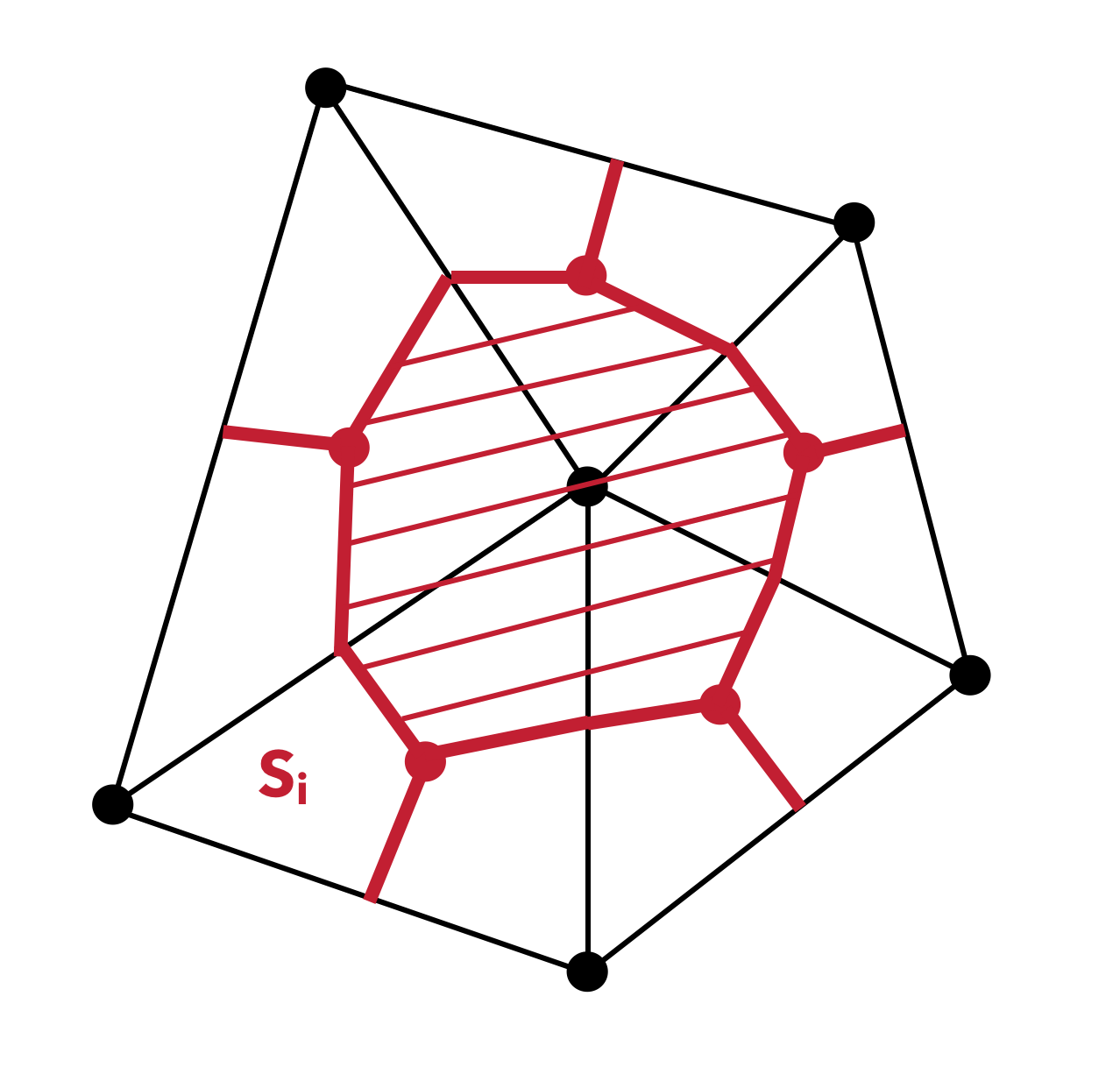}
    \caption{Dual cell (red shaded area) of a vertex in an unstructured triangular mesh}
    \label{fig:method_dual_cell}
\end{figure}

There are a number of important terms to define with respect to the domain discretisation and geometry. The 2D space $\Omega$ is completely divided into a set of triangular elements $T$, with vertices ($i$, $j$, $k$), labeled counterclockwise. The cells in the 1D case are now equivalent to the nodes of the triangulation, with the triangular elements taking the place of the cell boundaries. The inner normals of the triangle edges are defined such that $\textbf{n}_i$ is the inner normal to the edge between vertices $j$ and $k$. This is shown in Figure \ref{fig:method_normal}. These normals are
\begin{equation} \label{eq:method_normal}
    \textbf{n}_i = (y_j-y_k)\hat{\textbf{x}}-(x_j-x_k)\hat{\textbf{y}},
\end{equation}
where $\hat{\textbf{x}}$ and $\hat{\textbf{y}}$ are the $x$ and $y$ unit vectors respectively, with equivalent forms for the other vertices. It is also important to define the parameter $|S_i|$, representing the area of the dual cell of a vertex in a unstructured triangulation, shown in Figure \ref{fig:method_dual_cell}, given by
\begin{equation} \label{eq:method_dual_cell}
    |S_i| = \sum_{T|i \in T} \frac{1}{3}|T|,
\end{equation}
for the dual cell of vertex $i$, summing over every triangle $T$ with which $i$ is associated. $|T|$ is the area of the triangle, given by
\begin{equation} \label{eq:method_area}
    |T| = \frac{1}{2}|\textbf{n}_i\times\textbf{n}_j|,
\end{equation}
where $i$ and $j$ are any two vertices of $T$.
It is also important to define the specifics of the problem that is being solved. The system of partial differential equations depend on some set of continuous variables. For such a continuous variable $\theta(x,y,t)$, the equivalent discrete approximation is referred to as $\theta_h$. The parameter $h$ represents the characteristic length scale of an element, typically taken as the length of the longest edge.

\subsubsection{Residual}
For a set of PDEs with form as Equation (\ref{eq:pde_1D_first}), the linearised version of the problem becomes
\begin{equation} \label{eq:method_2D_jac}
    \frac{\partial \mathbf{Q}}{\partial t} + \mathbf{A_x} \frac{\partial \mathbf{Q}}{\partial x} + \mathbf{A_y} \frac{\partial \mathbf{Q}}{\partial y} = 0.
\end{equation}
The Jacobian is now a set of matrices $\mathcal{A}=(\mathbf{A}_x$, $\mathbf{A}_y)$, and as before these can be decomposed in such a way as to describe the problem as a set of characteristic waves. We will only describe the 2D case here, but will expand on the extension to 3D in Section \ref{sec:method_3D}.

The element residual, in this context, is now defined as the integral of the divergence of the numerical approximation of the flux \citep{ja:deconinck1993, ja:ricchiuto2010}. This follows naturally from the 1D definition, but the integral is now over the triangular element, rather that across the boundary between two cells. This is given as
\begin{equation}
    \phi^T(\mathbf{Q}_h) = \int_T \nabla \cdot \mathcal{F}_h(\mathbf{Q}_h) dx dy.
\end{equation}

The solution is found over a triangulation of the computational domain, built around a set of vertices, where the solution is calculated \citep{ja:ricchiuto2010}. The residual for each triangular element is divided between its vertices, with the sum of these residual from all the triangles of which a given node is a vertex, is the update to the state at the position of that node.

Defining the combination of the Jacobians as a single term $\mathbf{K}_i = (\bar{\mathbf{A}}_x n_{x,i}+ \bar{\mathbf{A}}_y n_{y,i})/2$, with the dependence on vertex $i$ coming from the normal of the opposite edge. This simplifies the calculation of the element residual to the sum of the product of matrix $\mathbf{K}_i$ and state $\mathbf{Q}_i$
\begin{equation} \label{eq:method_2D_res_final}
    \phi^T = \sum_{i=1}^3 \mathbf{K}_i \mathbf{Q}_i.
\end{equation}
A detailed summary of how this form is arrived at is given in Appendix \ref{app:RD_derivation_1st}. The residual is therefore split between the vertices of the triangular element, such that $\phi^T = \phi_i + \phi_j + \phi_k$. The discrete update to the state is found using the discretised form of the Taylor expansion of the solution. The discrete update becomes
\begin{equation} \label{eq:method_2D_update}
    \mathbf{Q}_i^n = \mathbf{Q}_i^{n+1} - \frac{\Delta t}{|S_i|} \sum_{T|i \in T}\phi_i
\end{equation}
where the summation is over all triangles for which node $i$ is a vertex. The area $|S_i|$ is the area associated with the updated vertex, defined by assigning one third of the area of each connected triangle. This area is also the area of the cell that is formed by the dual of the triangulation.

Second order accuracy in time is achieved using an explicit second order Runge-Kutta  (RK2) time-stepping algorithm \citep{ja:ricchiuto2010}. This approach makes use of an approximation of the time derivative of the residual, through the total residual $\Phi$. For the time step $n$ to $n+1$, the first step is constructed using the first order solver
\begin{equation}
    \mathbf{Q}_i^* = \mathbf{Q}_i^n - \frac{\Delta t}{|S_i|} \sum_{T|i \in T} \phi_i
\end{equation}
and the second step is found using the distribution of the total residual with
\begin{equation}
    \mathbf{Q}_i^{n+1} = \mathbf{Q}_i^* - \frac{\Delta t}{|S_i|}\sum_{T|i \in T} \Phi_i 
\end{equation}
where the total residual is calculated based on both the initial and intermediate element residuals, in the standard RK2 form \citep{ja:ricchiuto2010}.
The new residual term $\Phi_i$, known as the total residual, is given by
\begin{equation}
    \Phi_i = \sum_{j=1}^3 m_{ij} \frac{\mathbf{Q}^*_i - \mathbf{Q}^n_i}{\Delta t} + \frac{1}{2}\left(\phi_i(\mathbf{Q}^*_i) + \phi_i(\mathbf{Q}_i^n)\right),
\end{equation}
and estimates changes to the residual caused by the fluid state changing over the time step. As before, more detail is given in Appendix \ref{app:RD_derivation_2nd}. This approach is effectively a predictor-corrector setup, with the first order residual used as the predictor, and a correction applied that includes information about the time dependence of the solution. The exact form of the mass matrix term $m_{ij}$ is addressed in the following subsection.

These equations form the basis for the construction of a second order RD solver. The new total residual is only dependent on the initial state, the intermediate state, the time step, and the element residual for both the initial and intermediate states. However, it should be noted that second order accuracy is also dependent on the precise choice of distribution scheme, which is discussed below. We will demonstrate the second order accuracy numerically in Section \ref{sec:tests}.

\subsubsection{Distribution}
 We now describe how the residual is broken up between the constituent vertices of a given triangle, known as the \textit{distribution scheme}. When designing the distribution schemes, it is desirable that the scheme exhibit certain properties. Typically these include being conservative, and preserving linearity and positivity \citep{ja:deconinck1993, ja:ricchiuto2010, ja:abgrall2010}. A linearity preserving scheme will recover the exact solution for a linear set of equations, and a positive scheme will be total variation diminishing (TVD). For some set of initial conditions, a scheme is positive if it does not introduce new maxima or minima to the solution. It is impossible to construct a linear scheme that is both positive and linearity preserving \citep{ja:ricchiuto2010}, so non-linear approaches must be formulated to achieve both desired properties at once. This is essentially a statement of Godunov's theorem: linear numerical schemes that are monotone can be at most first-order accurate. To be clear, a linear \textit{scheme} mentioned here is one in which the solution can be expressed as a sum of the initial state, weighted by coefficients that do not depend on the state itself. This is not the same as the problem itself being linear. We cover three widely used schemes:
\begin{itemize}
    \item LDA Scheme - Linear, low diffusion
    \item N Scheme - Linear, positive
    \item B Scheme - Non-linear, blending of the two other schemes
\end{itemize}
A relatively simple example is the low diffusion A (LDA) scheme \citep{tb:struijs1995, ja:caraeni2002, ja:deconinck2007}. It achieves second order accuracy in space \citep{tb:struijs1995}, but sacrifices its total variation diminishing property. This results in spurious oscillations in the presence of discontinuities, such as shocks. It is constructed to have low numerical diffusion, making it effective for resolving smooth flows. The nodal residual, the part of the element residual sent to each vertex, is found using \citep{ja:csik2002}
\begin{equation}
    \phi_i^{LDA} = \beta_i \phi^T = \frac{\mathbf{K}_i^-}{\sum_{i}^3\mathbf{K}_i^-} \phi^T.
\end{equation}
To guarantee that this is conservative, it is only required that the distribution coefficients $\beta_i$ sum to unity, which ensures the distributed residual sum to element residual.

Another widely used method is the N scheme, which is designed to be positivity preserving, and does not experience the oscillations around discontinuities. The scheme is only first order accurate in space \citep{ja:ricchiuto2010}, so has greater numerical diffusion, compared to the LDA scheme. Such a scheme is best suited to problems with shocks. In this case, the nodal residual is given by \citep{tb:struijs1995}
\begin{equation}
    \phi_i^N = \mathbf{K}_i^+\left(\mathbf{Q}_i - \frac{\sum_{j=1}^3\mathbf{K}_j^-\mathbf{Q}_j}{\sum_{j=1}^3\mathbf{K}_j^-}\right).
\end{equation}
Both the above schemes are linear, and so cannot be both positivity and linearity preserving. As mentioned above, this leads to schemes that, in general, have either strong numerical diffusion, or weak shock handling capabilities. A number of non-linear schemes have been developed \citep{ja:csik2002, ja:abgrall2003, ja:dobes2008}, which build on these two linear schemes by combining them. The result preserves the advantages of each scheme, and reduces the disadvantages, by introducing a blending coefficient that can be designed to detect when the conditions are best suited to each method. A number of possible blendings have been developed, but they are built around the same idea of constructing the distribution around
\begin{equation}
    \phi_i^B = \Theta \phi_i^{N} + (I-\Theta)\phi_i^{LDA}
\end{equation}
where $I$ is the identity matrix, and $\Theta$ is the diagonal blending matrix. This matrix is constructed by setting
\begin{equation}
    \Theta_{ii} = \frac{|\phi^T_i|}{\sum_{j=1}^3|\phi^N_{j,i}|},
\end{equation}
where the sum is over every vertex of element $T$, and the $i$ index refers to the $i^{th}$ equation of the system. In this way, the change in each equation of the set is tested separately for the blending.

Different applications of this blending matrix put different conditions on the matrix values. The Bmax and Bmin schemes \citep{ja:csik2002, ja:paardekooper2017} simply replace every diagonal value of the blending matrix with either the maximum or minimum value of the blending matrix. Using Bmax will default towards the N scheme, while Bmin defaults to LDA. The so called Bx \citep{ja:abgrall2006,ja:dobes2008} scheme replaces the diagonal values with ones calculated with a shock sensor. This sensor detects when there are two colliding flows, where a shock will develop. Where these flows are detected, the N scheme will be heavily weighted. In all other conditions, the solution will use the LDA scheme. 

It is clear that the order of accuracy that can be achieved is not only dependent on the residual calculation itself, but is also dependent on the choice of distribution scheme. In particular, the schemes that produce a blending of other distributions have a potentially indeterminate accuracy. In these cases, and when we combine this problem with the fact of the linearisation of the underlying equations described in following sub-section, the most practical way to assess accuracy is to demonstrate the order through numerical tests, which we show in Section \ref{sec:tests}.



\subsection{Residual Distribution for the Euler Equations} \label{sec:method_euler}
As our implementation is aimed at eventually modelling the behaviour of baryonic gas, in an astrophysical context, we are specifically solving the Euler equations, which model the behavior of inviscid, compressible fluids. For this situation, the components of the 2D PDEs are given by
\begin{equation}
    \mathbf{Q} =
    \begin{pmatrix}
        \rho \\
        \rho v_x \\
        \rho v_y \\
        \rho E
    \end{pmatrix},\hspace{5mm}
    \mathbf{F}_x(\mathbf{Q}) =
    \begin{pmatrix}
        \rho v_x \\
        \rho v_x^2 + P \\
        \rho v_x v_y \\
        \rho v_x H
    \end{pmatrix}, \hspace{5mm}
    \mathbf{F}_y(\mathbf{Q}) =
    \begin{pmatrix}
        \rho v_y \\
        \rho v_x v_y \\
        \rho v_y^2 + P \\
        \rho v_y H
    \end{pmatrix},
\end{equation}
where the pressure $P$ is defined by the chosen equation of state. The other parameters have their usual physical meanings, $\rho$ is mass density, and $v_x$ and $v_y$ are $x$ and $y$ velocities respectively. These equations describe the conservation of mass, momentum and energy. We use the equation of state 
\begin{equation}
    P = \rho(\gamma - 1)\left(E - \frac{\textbf{v}\cdot\textbf{v}}{2}\right).
\end{equation}
The other variables are defined as normal, where $\rho$ is mass density, $\mathbf{v} = (v_x, v_y)$ are the velocity components, and $E$ is the specific energy, with enthalpy $H$ given by
\begin{equation}
    H = E + \frac{P}{\rho}.
\end{equation}
The adiabatic speed of sound $c_s$ is defined as
\begin{equation} \label{eq:sound_speed}
    c_s = \sqrt[]{\frac{\gamma P}{\rho}}.
\end{equation}
By definition these equations assume the fluid has no viscosity. As mentioned before, this assumption is applicable to many astrophysical scenarios, where viscosity appears to be negligible. Equivalent RD methods have been developed for the viscous Navier-Stokes equations \citep{ja:abgrall2015}.

\subsubsection{Roe Parameters}
For the Euler equations, Roe produces a parameter vector that can suitably linearise the Euler equations \citep{ja:roe1981}. For the standard fluid variables given above, the Roe parameter vector is $\mathbf{Z}=(\sqrt{\rho},\sqrt{\rho}v_x,\sqrt{\rho}v_y,\sqrt{\rho}H)^T$. The Eulerian flux vector has a quadratic dependence on these variables, which means that the Jacobian $\mathcal{A}$ is now linearly dependent on the state variables. For a detailed summary of how the Euler equations are linearised and combined with the RD approach, see Appendix \ref{app:RD_derivation_Euler}.

\subsubsection{Time Step}
To calculate the update using Equation (\ref{eq:method_2D_update}), a mechanism to calculate a suitable time step is required. As has been discussed previously, the time step choice is not arbitrary. The CFL condition, that the numerical domain of dependence should enclose the physical domain of dependence, fundamentally limits the time steps that will produce a physically accurate result. Such a condition can be achieved \citep{ja:ricchiuto2010} by requiring that the time step $\Delta t$ is limited by
\begin{equation} \label{eq:method_euler_2D_timestep}
    \Delta t \leq \min_{i \in \mathcal{T}}\frac{2|S_i|}{\sum_{T|i \in T} l_{\mathrm{max}}^T \lambda_{\mathrm{max}}^T},
\end{equation}
where $l_{\mathrm{max}}^T$ is the longest edge of triangular element $T$, and $\lambda_{\mathrm{max}}^T$ is a measure of the maximum speed at which information can move across the element. This is done by setting
\begin{equation}
    \lambda_{\mathrm{max}}^T = \max_{j \in T} (|\mathbf{v}_j| + c_j).
\end{equation}
This is the maximum of the combination of the fluid speed and sound speed at the vertices of the triangle, which is equivalent to the maximum signal speed in that element. Together, the product of this length and signal speed, multiplied by a factor of a half, produce an estimate of the area per time of an imaginary triangle swept out by the material in this element. Summing up the contributions from all the element associated with a vertex $i$, and dividing the actual area associated with that vertex by this value, produces an estimate of the time it will take a signal to propagate across the dual cell. Keeping the time step below the minimum such value required by any vertex in the mesh $\mathcal{T}$ produces a limit within which the CFL condition will always be met. In practice, some fraction of this value will be usually be used, as an additional guarantee that the condition is not breached. This fraction typically varies between $0.1$ and $0.5$, depending on the complexity of the problem.

\subsubsection{Summary of Equations}

In the previous sections, we cover the theoretical background to the RD solvers development and extension. We also describe the precise formulation required to construct an RD solver for the 2D Euler equations. We now briefly summarise the most important equations. The update to the fluid state $\mathbf{Q}_i$ at vertex $i$ , from time step $n$ to $n+1$, is given by
\begin{equation} \label{eq:method_euler_sum_intup}
    \mathbf{Q}_i^{*} = \mathbf{Q}_i^n - \frac{\Delta t}{|S_i|} \sum_{T|i \in T} \phi_i,
\end{equation}
and
\begin{equation} \label{eq:methodeuler_sum_secup}
    \mathbf{Q}_i^{n+1} = \mathbf{Q}_i^* - \frac{\Delta t}{|S_i|} \sum_{T|i \in T} \Phi_i,
\end{equation}
where the time step is given by Equation (\ref{eq:method_euler_2D_timestep}), and the dual area is
\begin{equation} \label{eq:chaprd_euler_sum_dual}
    |S_i| = \sum_{T|i \in T} \frac{1}{3}|T|.
\end{equation}
The nodal residual is calculated from the element residual, based on the chosen scheme, and the element residual
\begin{equation} \label{eq:method_euler_sum_elemres}
    \phi^T = \sum_{j=1}^3\mathbf{K}_i\hat{\mathbf{Q}}_i,
\end{equation}
where $\hat{\mathbf{Q}}_i$ is the linearised fluid state (see Appendix \ref{app:RD_derivation_Euler}). The exact form of $\mathbf{K}_i$ can be found in the appendix of \citet{ja:paardekooper2017}. The equivalent total residual is calculated using the element residual
\begin{equation} \label{eq:method_euler_sum_totres}
    \Phi^T = \sum_{j \in T} m_{ij}\frac{\mathbf{Q}_j^*-\mathbf{Q}_j^n}{\Delta t} + \frac{1}{2}\left(\phi_i(\mathbf{Q}^*) + \phi_i(\mathbf{Q}^n)\right).
\end{equation}
The mass matrix form varies with the scheme, as does the distribution itself. The linearised state $\hat{\mathbf{Q}}_i$ is calculated with
\begin{equation} \label{eq:method_euler_sum_qhat}
    \hat{\mathbf{Q}}_i = 
    \left(
        \begin{matrix}
            2 \bar{Z}_1 Z_1 \\
            \bar{Z}_2 Z_1 + \bar{Z}_1 Z_2 \\
            \bar{Z}_3 Z_1 + \bar{Z}_1 Z_3 \\
            \frac{1}{\gamma}\left(\bar{Z}_4 Z_1 + \gamma_1 \bar{Z}_2 Z_2 + \gamma_1 \bar{Z}_3 Z_3 + \bar{Z}_1 Z_4 \right)
        \end{matrix}
    \right).
\end{equation}
Together these equations describe all the key variables and functions needed to construct the 2D RD hydro-solver. Combining these with the distribution schemes described above, one can produce a fully functioning RD solver.

\subsection{3D Extension} \label{sec:method_3D}
Some astrophysical systems can be effectively modelled using only two dimensions, such as thin discs, where useful results can be found without calculating flows in the third direction. However, many more systems, from the cosmic web down to giant molecular clouds, are more accurately described by the full three dimensional flows. The RD approach naturally extends to extra dimensions, as the basic form is generalised to any number of dimensions \citep{ja:paardekooper2017}. The set of PDEs is now simply the standard 3D form of the Euler equations. For the sake of clarity, we will briefly cover the fundamental changes to the underlying geometric and physical definitions.

All summations over the three vertices of a triangular element simply become summations over the four vertices of the 3D simplex, a tetrahedron. Equivalently, the dual area associated with each vertex is now a dual volume, and the normals associated with each vertex are now the inward facing normals to the opposite face, rather than opposite edge. All vector quantities have an additional $z$-component. Most notably the key problem that must be solved is for the form of the inflow matrix $\bm{K}_i$, and the associated $\bm{K}_i^-$ and $\bm{K}_i^+$. These matrices contribute directly to the calculation of the element residual, given in Equation (\ref{eq:method_euler_2D_res_final}), and to the different distribution schemes. Their definition depends on their Schur decomposition form. The exact form of this decomposed version are given in Appendix \ref{app:3D_kmatrix}.

\section{Mesh} \label{sec:triangulation}
The RD approach is built around an arbitrary set of static tracer positions $\mathbf{r}$, or vertices, which require a set of simplices that fill the periodic domain, without gaps, and without overlapping edges/faces. In two dimensions, these simplices are triangles. A given distribution of vertices can have a large number of possible meshes that fulfill the above criteria, but many of these will have undesirable characteristics. For our purposes, we require a triangulation that minimises the existence of extremely elongated triangles. The more elongated the triangles, the shorter the time-step required by Equation (\ref{eq:method_euler_2D_timestep}), and so the more computationally expensive finding the solution becomes. The triangulation itself should also be as computationally cheap as possible to construct. While not currently included in the code, on the fly re-triangulation of the domain, either to regularise the vertex distribution, or to refine the spatial resolution in areas of interest, could be valuable additions to the code. Choosing an approach that plans for these possibilities is therefore important.

Independent of the precise triangulation that is chosen, the natural construction of the RD approach around an arbitrary set of vertex positions is a core advantage of this method. Unstructured meshes are particularly useful in capturing the highly unstructured flows found in astrophysical scenarios, such as the turbulent and clumpy interstellar medium. They avoid preferential flow directions, minimising the existence of spurious features. The natural construction around these meshes is therefore a advantageous characteristic.

\subsection{Delaunay Mesh} \label{sec:tri_del}
In this work, the discretisation of the gas is built around a Delaunay triangulation. This triangulation maximises the minimum opening angle of the underlying triangles. There are a number of well documented methods of constructing the Delaunay triangulation. These methods, and the corresponding numerical algorithms, have been extensively tested, resulting in extensive, highly optimised, construction libraries \citep{ja:springel2009, tb:cheng2012, ja:duffell2016}. The Delaunay triangulation is defined by the Delaunay condition

\begin{theorem} \label{def:delaunay}
(Delaunay condition) In the context of a finite point set S, a Delaunay triangle is characterized by the empty circumdisk property: no point in S lies in the interior of any triangle's open circumscribing disk in 2D, or circumscribing sphere in 3D.
\end{theorem}

\begin{figure}
    \centering
    \includegraphics[width=0.45\textwidth]{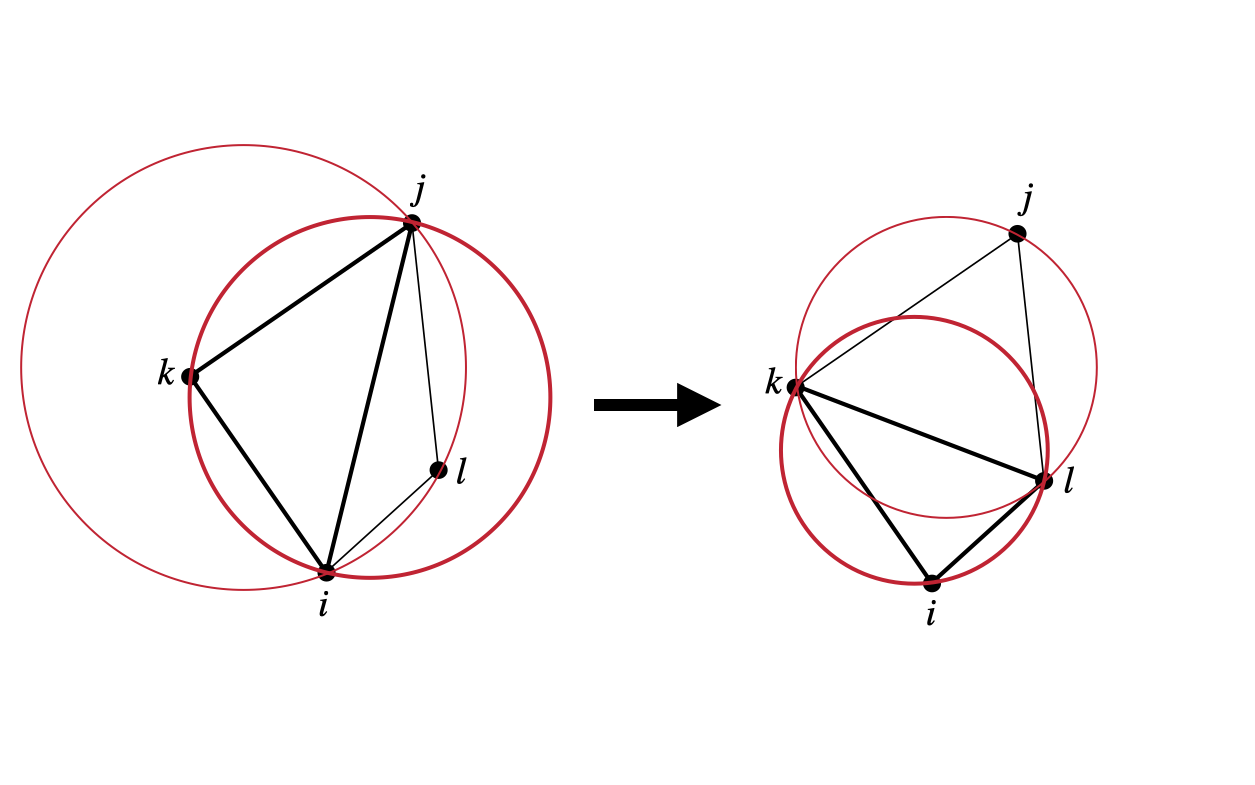}
    \caption{Edge flip process. Moving edge $ij$ to $lk$ changes the two non-empty circumdisks of triangles $ijk$ and $ilj$ into two triangles ($ilk$ and $ljk$) with empty circumdisks. The two triangles on the left are therefore non-Delaunay, while the two on the left satisfy the Delaunay condition.}
    \label{fig:triang_edge_flip}
\end{figure}

The left hand side of Figure \ref{fig:triang_edge_flip} shows an example of a circumdisk for triangle $ijk$ as the thick red line. If all simplices within a given triangulation satisfy the Delaunay condition, then the triangulation is the Delaunay triangulation, which is unique for the vast majority of vertex distributions \citep{tb:cheng2012}. The only major exceptions are some ordered distributions, which typically have two triangulations that satisfy a relaxed version of the condition, where only the closed circumdisk must be empty. In practical terms, in these scenarios, it is acceptable to simply pick either one of the options.

An example of a non-empty circumdisk is shown on the left hand side of Figure \ref{fig:triang_edge_flip}. The red circles show the circumdisks for $ijk$ and $ilj$, both of which enclose vertices that are not part of that triangle. On the right hand side, we show two examples of triangles with empty circumdisks. The circumdisks of triangles $ilk$ and $ljk$ do not contain any external vertices, and so these triangles satisfy the Delaunay condition. The Delaunay triangulation will exist uniquely for any set of points, except in a small set of scenarios where the open circumdisk is not empty because of an additional vertex just on the edge of the disk \citep{tb:cheng2012}. In these cases the Delaunay condition can be relaxes to only require the closed circumdisk to be empty.

There are several characteristics of the Delaunay triangulation that make it appealing for fluid discretisation. The Delaunay condition of requiring an empty circumdisk innately maximises the minimum opening angle, and also minimizes the largest circumdisk. This has the effect of minimizing distortion, but specifically for cells with obtuse angles. Very large or very small angles may still exist in an arbitrary point distribution, but will be minimised by this setup. To further reduce distortion, methods for shifting the vertices of the tessellation, without reducing the physical accuracy of the simulation, have been proposed \citep{ja:springel2009}. Such approaches attempt to regularise the distribution of vertices.

\subsubsection{Mesh Construction} \label{sec:tri_CGAL}
Given some arbitrary distribution of vertices, a Delaunay triangulation can be constructed by an number of different approaches. These include: \textit{edge flipping}, where the edges of an initial arbitrary triangulation are flipped in turn, producing Delaunay triangles (process shown in Figure \ref{fig:triang_edge_flip}); \textit{gift wrapping}, where new Delaunay triangles crystallise around a known Delaunay triangulation; and \textit{incremental insertion}, where vertices are added one at a time, weighted to insert vertices that are closest to current edges first.

In practice, the various construction mechanisms are used in combination to produce the most efficient algorithm. A number of extensive C/C++ libraries have been developed to produce periodic Delaunay triangulations in both 2D and 3D. We have integrated triangulation construction from the CGAL (https://www.cgal.org/) library, due to its inclusion of 3D periodic triangulations. These libraries efficiently construct a periodic Delaunay triangulation for any distribution of vertices, and include handling of the rare vertex sets with non-unique Delaunay triangulations. Any example of this scenario is a Cartesian cubic grid of vertices. In these cases, the CGAL functions systematically select one of the possible orientations, avoiding locking up in infinite loops of edge flipping, for example.

\section{Adaptive Time-Stepping} \label{sec:timestep}
As discussed before, the time step is defined by the CFL condition, requiring that the numerical domain of dependence encloses the entire physical domain of dependence. In the standard formulation, the time step at which the fluid state at every vertex is updated, is dictated by the smallest time step required by any vertex in the mesh. In scenarios with regularised meshes, or with only small variations in density and velocity across the grid, this implementation is not particularly significant, as all vertices will require similar time steps. However, scenarios with extreme density and velocity contrasts will be computationally inefficient to run. If only a few cells require a time step that is orders of magnitude shorter than the rest of the mesh, then the residuals will be recalculated far more often than is numerically required for many triangles. This problem is encountered in all numerical methods for solving fluid dynamics problems. There is always some limit on the time step at which the cells or particles can be updated, and large variations in this number lead to inevitable inefficiencies. To combat this problem, and produce methods that can utilise the available computational power more efficiently, many numerical methods have introduced mechanisms that allow different fluid elements to be updated with different time steps.

Typically, these approaches divide elements into groups based on their required time step, and then recalculate the evolution of the state based on these time step bins. For simplicity we will only discuss grid based approaches from here, but equivalent particle methods are widely used. As discussed in detail in the introduction, in a standard cell based approach (for either a structured static mesh, or an unstructured or even moving mesh), the fluid state is updated by calculating the flux of material through the faces of cells. A simple way to implement a varied time step in such methods is outlined in \citep{ja:springel2009}. You first bin each cell by the required time step, then to recalculate the flux through faces based on the smallest value either side of that face. The fluid state is still updated at the smallest time step of the whole mesh, but the flux is not recalculated for every face at every small time step. The flux through faces whose required time step is longer than this is simply kept the same, until it is necessary to update it. Using old updates is known as drifting, as the state continues on the same trajectory for multiple small time steps.

\begin{figure}
    \centering
    \includegraphics[width=0.45\textwidth]{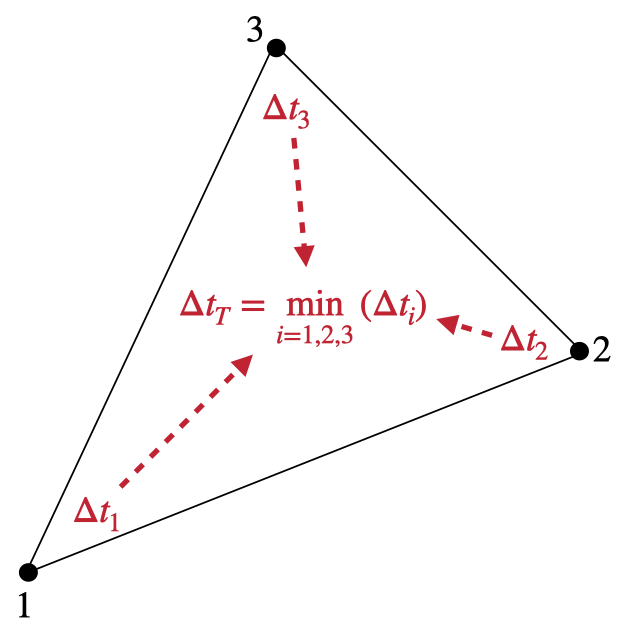}
    \caption{The time step bin of the triangle $\Delta t_T$ is the minimum of the bins assigned to the vertices of the triangle ($\Delta t_1$,$\Delta t_2$,$\Delta t_3$).}
    \label{fig:chaprdext_time_tri}
\end{figure}

To produce the equivalent effect with the RD solver, it is the residual that is calculated at different steps for different triangles, as this is the analogous calculation to the flux in the standard grid approach. We have implemented a novel strategy to achieve this outcome. The minimum time step for every \textit{vertex} is calculated using the limit described in the previous chapter. Each \textit{triangle} is then binned based on the smallest time step required by any of its vertices. The time step bins have limits based on powers of two times the overall minimum time step $\Delta t_\mathrm{min}$, such that the smallest bin has limits $\Delta t_\mathrm{min} < \Delta t_\mathrm{req} < 2\Delta t_\mathrm{min}$, the next bin has $2\Delta t_\mathrm{min} < \Delta t_\mathrm{req} < 4\Delta t_\mathrm{min}$, and so on. Now when the simulation is evolved, residuals are only recalculated after the lower limit of their time step bin has elapsed since last calculation. Every triangle is checked, but only some have their residual recalculated, which are referred to as \textit{active} triangles. This saves significant computational time by not recalculating residuals more frequently than required. Over the course of the large time step, which is defined as the lower limit of whichever is the largest time step bin, the following will happen. Taking the simplest example of only two time step bins, there will be two small time steps modelled for the one large time step. At the beginning of the large time step, the residual of every triangle is calculated, but one small time step later, only the triangles in the smallest bin will recalculate their residual. Another small time step later, we have completed a full top level time step, and so start the process over. The passing of the update is described below. Once the top level time step has been completed, the binning process is repeated.

\begin{figure}
    \centering
    \includegraphics[width=0.45\textwidth]{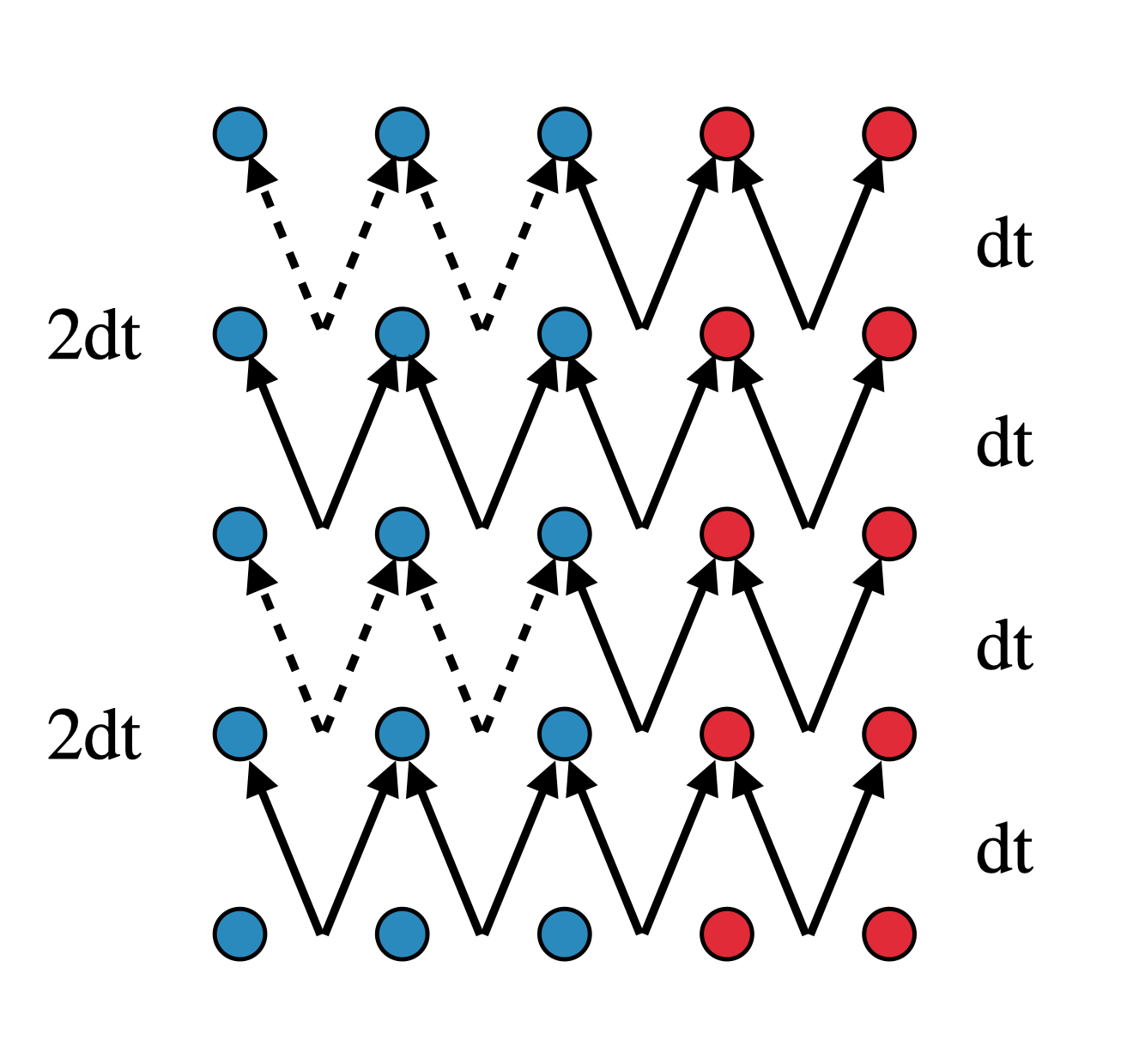}
    \caption{Stencil for the \texttt{DRIFT} adaptive showing distribution of residual in 1D, with time increasing in the $y$-direction. Dots represent the vertices where the fluid state is held, while the spaces between them in the $x$-direction are the elements for which the residuals are calculated. Red vertices require time steps of dt, and the blue dots 2dt. The arrows show where residuals are distributed. The solid arrows represents residuals that have been recalculated that turn, while the dashed line represents residuals that have not been updated.}
    \label{fig:chaprdext_stencil_drift}
\end{figure}

The update is passed based on the current residual of that triangle, even if that residual has not been recently recalculated. The updates from long time step bins can be said to drift the state of the associated vertices because the changes continue along constant trajectories, as if they are drifting in some direction, without being deflected by additional forces, hence the approach's label as the \texttt{DRIFT} method. A 1D representation of the concept of this method is shown in Figure \ref{fig:chaprdext_stencil_drift}. The red discs represent vertices that are in the bottom level, shortest time step bin, while the blue discs represent those in the top level bin (assuming a two bin system). The spaces between each disc represents the element for which the residual is calculated. Time increases in the $y$-direction, with each row representing the vertices at a given time. The arrow represents the passing of an update, based on the element's residual, to each associated vertex. Solid arrows show the distribution of residual calculated that time step, while dashed arrows show the passing of residual calculated at a previous time. For the first small time step dt, the residual passed to each triangle is exactly the same as it would be in the original system, but for the second small time step, the residual from the left two elements are based on an old fluid state. They have not been updated, as the fluid states at the vertices of these elements do not require the short time steps.

We find that this approach provides a significant boost to performance, but also that there is a loss of exact conservation. The total mass and energy of the box change over time, while for the basic universal time step method preserves conservation to machine precision. The loss of conservation is caused by the peculiarities of the residual method itself. In typical grid based methods, where the change in the fluid state of a cell is calculated by estimating fluxes through the surface of the cell, conservation is trivial to maintain for most situations. Any material that flows from one cell is added to its neighbour. Conservation is explicitly maintained across every cell boundary. The residual method, however, does not maintain conservation across the equivalent structures: the triangular elements. For a given element the net change in the states of all vertices, produced by the residual, is not zero. It is zero in the special case of a steady flow, when the element residual is also zero. Conservation is ensured between a vertex, and all of its neighbours, but only once all residuals have been distributed. In the normal, global time step, setup, this is perfectly adequate, as all residual are recalculated every time step, and so consistent updates are present everywhere.

However, when we use the \texttt{DRIFT} approach, some updates are based on outdated residuals. These updates assume vertex states that no longer exist. In the standard flux approach, this is not a problem for conservation, because even if the current flux is not exactly physically correct, it is the same incorrect value on either side of the face. In the RD case, neighbouring triangles use residuals from inconsistent states, effectively breaking the guarantee of conservation, which relies on neighbouring triangles calculating residuals from the same states at the shared vertices.

\begin{figure}
    \centering
    \begin{tabular}{c}
        \includegraphics[width=0.45\textwidth]{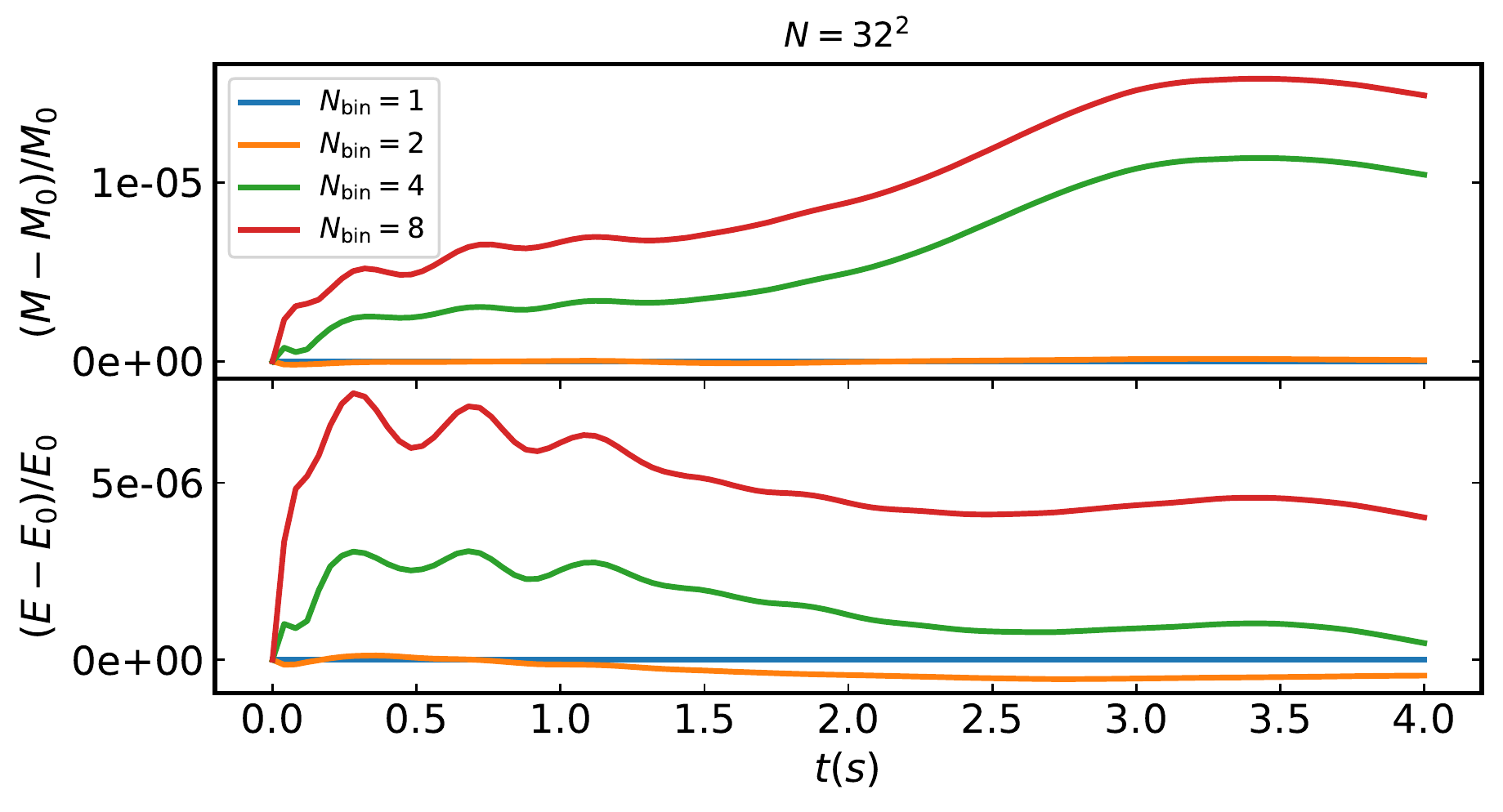}  \\
        \includegraphics[width=0.45\textwidth]{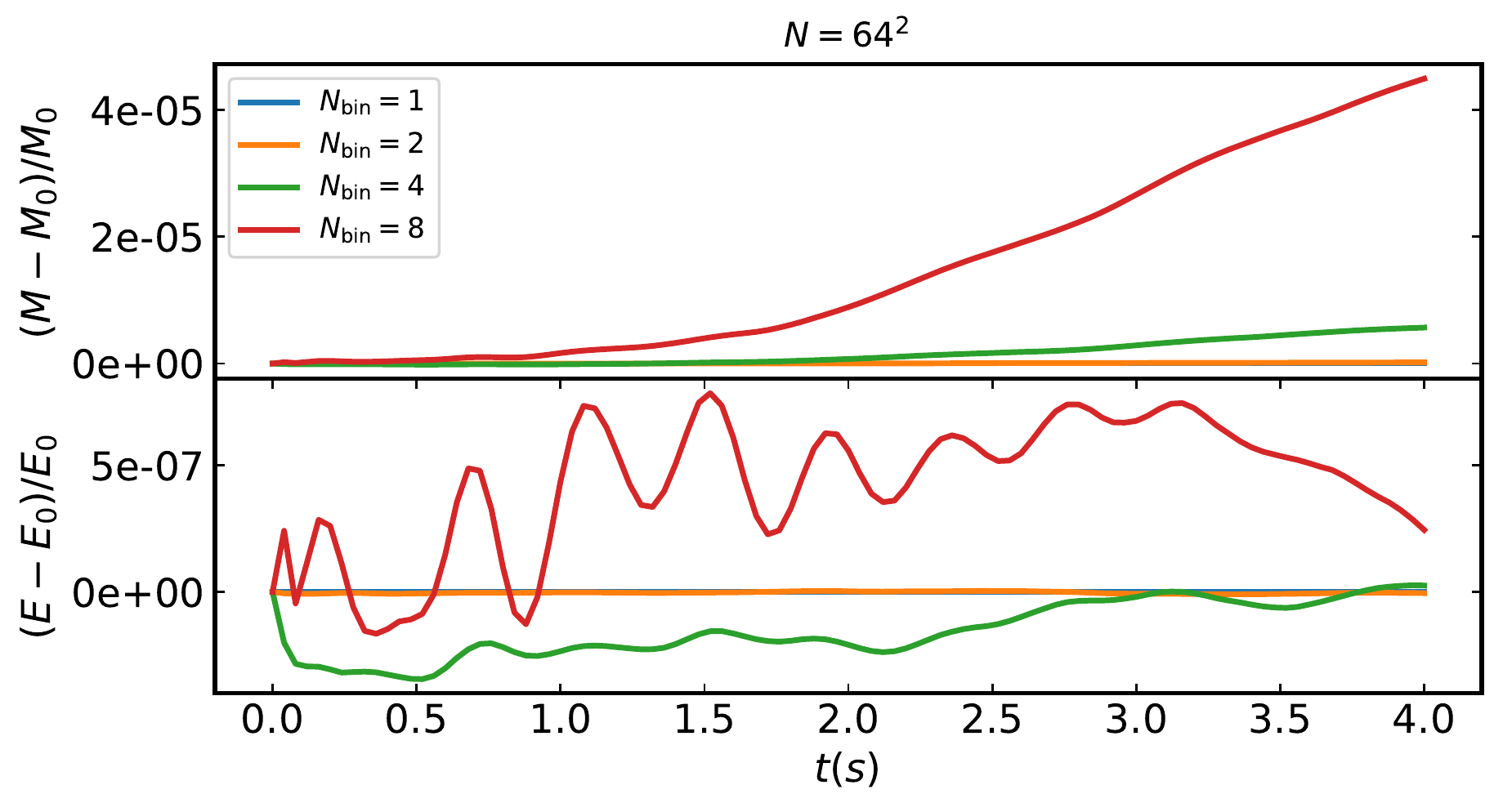}  \\
        \includegraphics[width=0.45\textwidth]{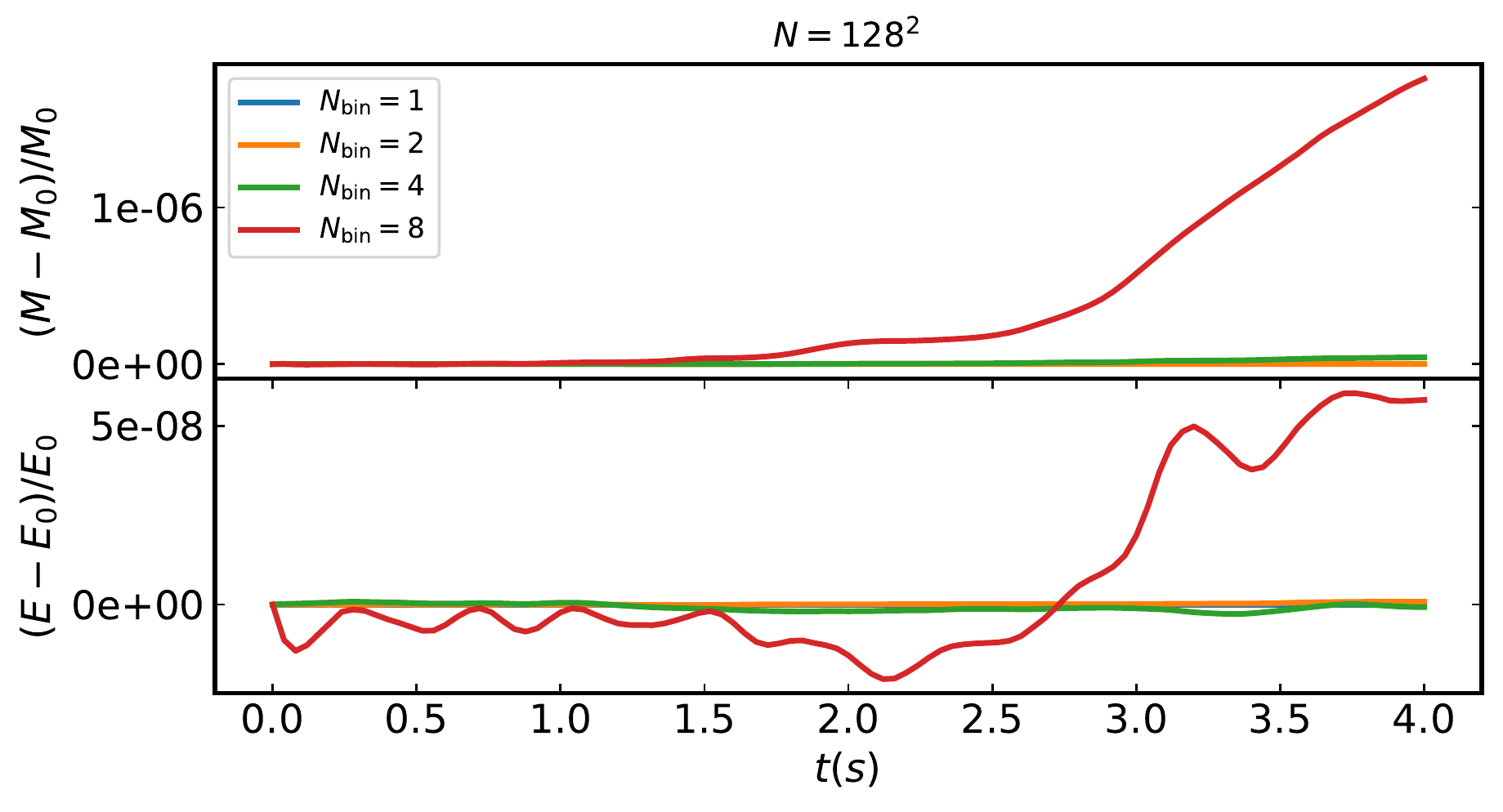}
    \end{tabular}
    \caption{Variation in total mass (upper part of each panel) and energy (lower part of each panel) using the \texttt{DRIFT} method, for $N=32^2$ (top panel), $N=64^2$ (middle panel), and $N=128^2$ (bottom panel). These show the fractional change from the initial total mass and energy. The lines show the results for different numbers of time step bins, where we have $N_\mathrm{bin}=1$ (blue), $N_\mathrm{bin}=2$ (orange), $N_\mathrm{bin}=4$ green), and $N_\mathrm{bin}=8$ (red).}
    \label{fig:time_con}
\end{figure}

In Figure \ref{fig:time_con}, we show the evolution of total mass and energy in the Kelvin-Helmholtz tests (see Section \ref{sec:tests_2D}), which show the impact of this potential loss of conservation. Each plot shows the difference between the current total mass, or energy, and the initial value, as a fraction of this initial value. From top to bottom, we show the results for $N=32^2$, $64^2$, and $128^2$, including lines for different numbers of time-step bin, $N_\mathrm{bin}=1$ (blue), $N_\mathrm{bin}=2$ (orange), $N_\mathrm{bin}=4$ (green), and $N_\mathrm{bin}=8$ (red). The larger the number of time-step bins, the greater the loss of conservation. This is to be expected, as the larger the number of bins, the greater the number of vertices with neighbours that are in different bins. The change in both mass and energy does decrease with increased resolution, which is desired, as the adaptive time-stepping regime is of most use with very large problem sets. The key take-away from these plots is that even when present the total error over many thousands of time-steps, as shown in these plots, is still only at the fraction of a percent level, allowing us to use the method with confidence.

\section{Tests} \label{sec:tests}

A wide range of standard hydrodynamics tests exist that allow us to compare and quantify the abilities of numerical hydrodynamics methods. We focus on a number of tests, which demonstrate the multidimensional advantages of the residual distribution approach. This includes tests run in one, two and three dimensions, and comparison of our numerical solution to both analytic solutions and numerical solutions from other solvers used in current astrophysical simulations. All initial conditions for these tests are generated by test cases within the code, with the specific setups for each tests described below. The vertices, around which the Delaunay mesh is generated, are distributed either randomly, or in a structured distribution that will produce a mesh of uniform triangles. Whenever we refer to a uniform mesh, we are referring to a set of vertices that are created by taking vertices placed on a Cartesian grid, and offsetting every second row by half the vertex separation distance. The chosen distribution is specified for each test.

\subsection{1D Test}
We run a standard one dimensional test. This has a well defined analytic solutions, so poses a good initial test of the solver's ability to capture both advection and shocks. It is run in pseudo 1D (i.e. they are run on 2D meshes, but with no variation in one direction). By doing this, we effectively also check for spurious flows/dissipation in the extra dimension where, physically, nothing should be changing.

\subsubsection{Sod Shock Tube}
Shocks are a common feature of many astrophysical systems, found in cosmic filaments, gas falling into dark matter halos, the formation of stars, and the supernovae at the end of stellar lifetimes. The Sod shock tube \citep{ja:sod1978} sets up a simple 1D shock, with a well defined solution to the evolution of the density, velocity, and pressure.

\begin{figure}
    \centering
    \includegraphics[width=0.4\textwidth]{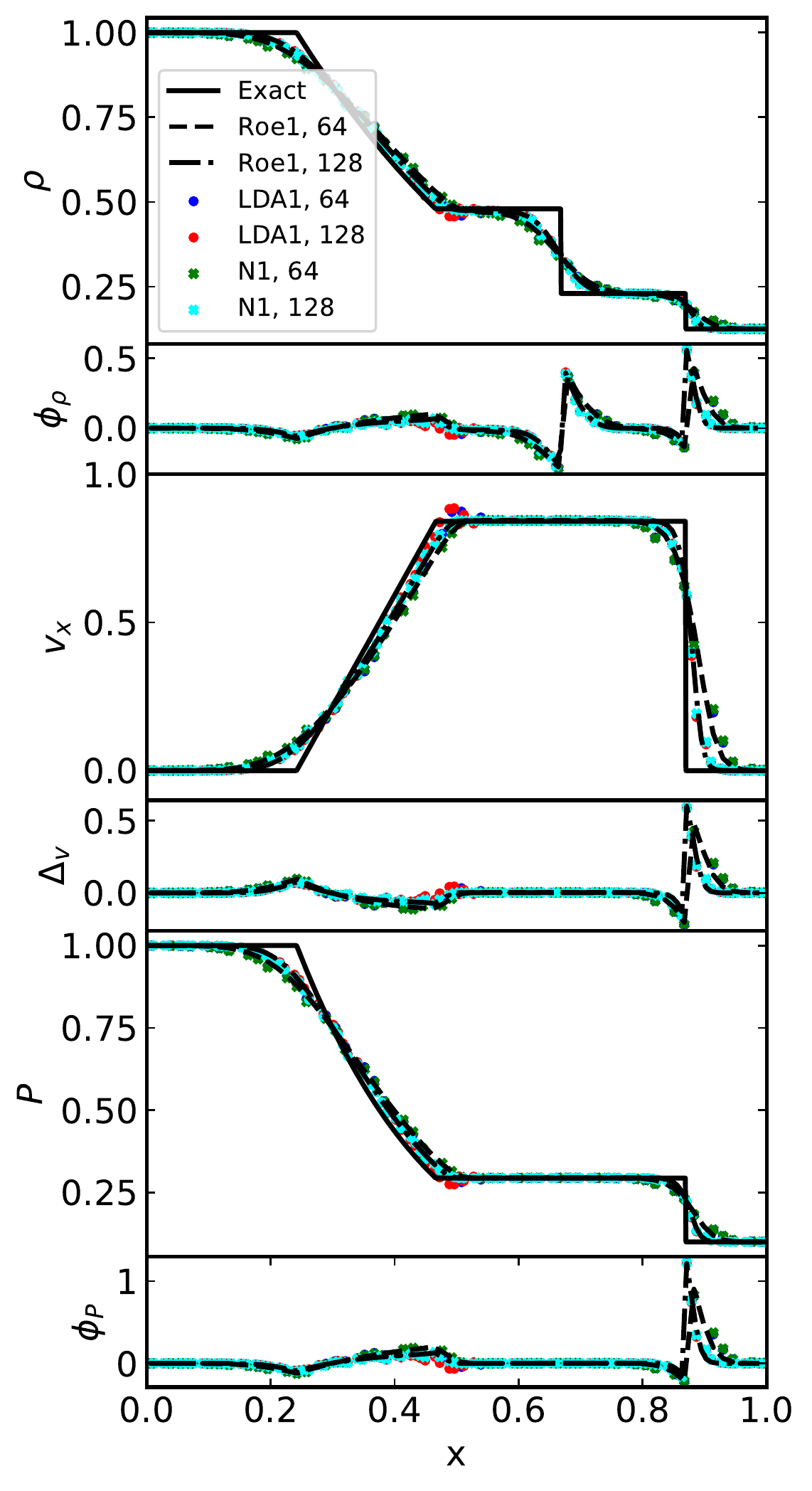}
    \caption{Sod shock tube for the first order LDA and N schemes (in pseudo 1D), and the Roe solver (in true 1D). $N=64$ and $N=128$ vertices in the $x$-direction. Dots show LDA results (blue for $N=64$ and red for $N=128$), and crosses N scheme results (green $N=64$ and cyan $N=128$). Solid black line is the exact solution. From top to bottom, major panels show density, $x$-velocity and pressure, with the minor panels showing the difference between the numerical and exact solutions $\phi_X=(X_\mathrm{num}-X_\mathrm{exe})/X_\mathrm{exe}$, for each property, apart from those tht have significant regions where the analytic solution is zero, where we show the difference $\Delta_X=X_\mathrm{num}-X_\mathrm{exe}$.}
    \label{fig:tests_sod}
\end{figure}

The initial conditions consist of two regions, each that fill half of the box. The velocity is zero everywhere. The left hand side of the tube has density $\rho_L=1$ and pressure $p_L=1$, with the right hand side of the tube at $\rho_R=0.125$ and $p_R=0.1$. These are run using $\gamma = 5/3$, and CFL coefficient of 0.4, meaning the time step is two fifths of the value strictly required by the time step condition. This test uses the uniform vertex distribution. Figure \ref{fig:tests_sod} shows the results from both the first order LDA and N schemes, compared to the exact solution for the Sod shock tube. We have also included results from the 1D Roe solver. The LDA results are shown as dots, with $N=64$ vertices in the $x$-direction in blue, and $N=128$ in red. The N scheme results are represented by crosses, in green ($N=64$) and cyan ($N=128$). The black solid lines give the exact solution, while the dashed lines show the results for the first order Roe solver, for $N=64$ and $N=128$ respectively. Below each panel, we show the difference between the numerical and analytic solutions. The results from the Roe solver closely match the results from the RD solvers, as expected. The minimal differences are caused by the Roe solver being applied to truly 1D grid, whereas the RD solvers are run on a pseudo 1D mesh, where there is no variation in the $y$-direction. There is a small amount of numerical dissipation from material flowing in the $y$-direction, even though the resultant variation in that dimension is zero.

There is clear evidence of spurious oscillations at $x=0.5$ for the LDA1 solver, as predicted, but it decreases with better spatial resolution. The N scheme does not show these structures, as expected, but the profiles at the transitions between solution phases do show smoothing. This is present in both LDA and N scheme solvers, as well as in the Roe solver solution. The smoothing is improved in all cases by the increase in resolution, and is caused by the numerical diffusion inherent in the method. Further increasing the resolution will improve the sharpness of these profiles.



\subsection{2D Tests} \label{sec:tests_2D}
The tests discussed so far demonstrate how well the RD solvers recover solutions that are well known for 1D flows, where the exact solution can be known. The key difference that this RD approach has, when compared to the standard mesh based methods of most approaches currently used in the field, is the truly multi-dimensional way in which the equations are solved. With the 2D tests discussed in this section, it is possible to demonstrate the ability of this solver to handle complex multi-dimensional flows.

\subsubsection{Kelvin-Helmholtz Instability}
Kelvin-Helmholtz instabilities form at the interface between shear flows. These occur in terrestrial and astrophysical contexts, such as between cloud layers in our atmosphere, or, at the other end of the size scale, in jets from AGN. This test sets up such a scenario, with two regions of gas moving alongside each other in opposite directions. The periodic box of side length $L_x = L_y = 1$ has a central region, with boundaries $0.25<y<0.75$, with density $\rho_0 = 2$, moving in the $x$-direction with velocity $v_{x0}=1$. The outer region is moving with velocity $v_{x1}=-1$, and has density $\rho_1=1$. The vertex distribution is uniform. The difference in density is not important for the instability itself, but is useful in observing the mixing of the two flow. To generate the instability in a systematic way, a very small transverse velocity is introduced, with sinusoidal variation in the direction of the flow. The instability is expected to develop into a spiral like structure, as the two flow mix at the boundary. The main quantitative test of the results is to compare the growth of transverse kinetic energy. This can only be done while the instability remains linear. Other than that, qualitative comparisons are limited to the sharpness of the boundary between density components, as a measure again of numerical diffusion.

\begin{figure}
    \centering
    \includegraphics[width=0.45\textwidth]{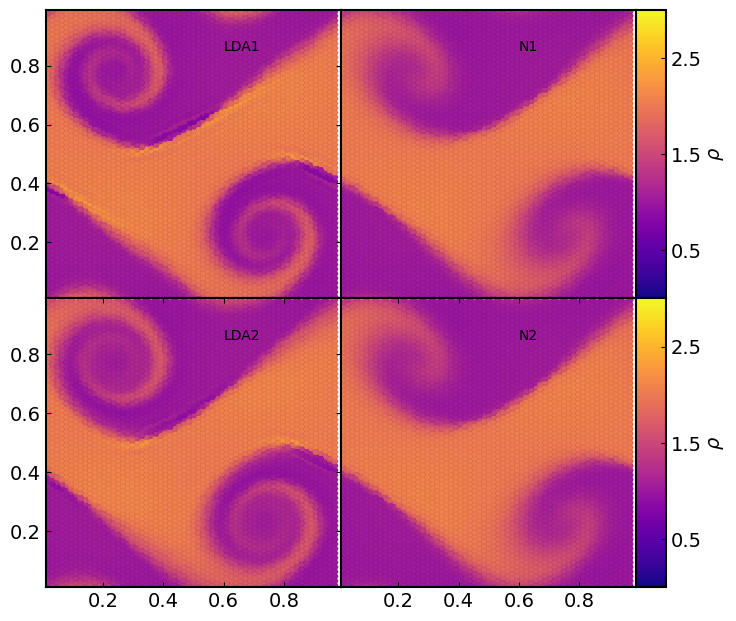}
    \caption{Kelvin-Helmholtz instability for the first (top row) and second (bottom row) order LDA (left column) and N scheme (right column) solvers. The color scale shows variation in density. All cases use $N=64^2$.}
    \label{fig:tests_khy64}
\end{figure}

The density results of evolving this shear flow setup for the various solvers are shown in Figure \ref{fig:tests_khy64} for $N=64^2$, with LDA1 (top left), LDA2 (bottom left), N1 (top right), and N2 (bottom right). The expected structures form very clearly in the LDA cases, with the winding structure recovered down to a few cells across. All KH plots show results at $t=2$ unless otherwise stated. The N scheme results show much less structure, with only the broad curling of the flow being recovered. This is caused by the numerical diffusion of the scheme, which the second order formulation does not significantly change. The LDA results can resolve the structure in the greatest detail, and so are favourable for problems that involve complicated flows, without any shocks. Running the same test with a blended scheme largely reproduces the LDA results, as the blending favours those residuals in these conditions.

\begin{figure}
    \centering
    \includegraphics[width=0.4\textwidth]{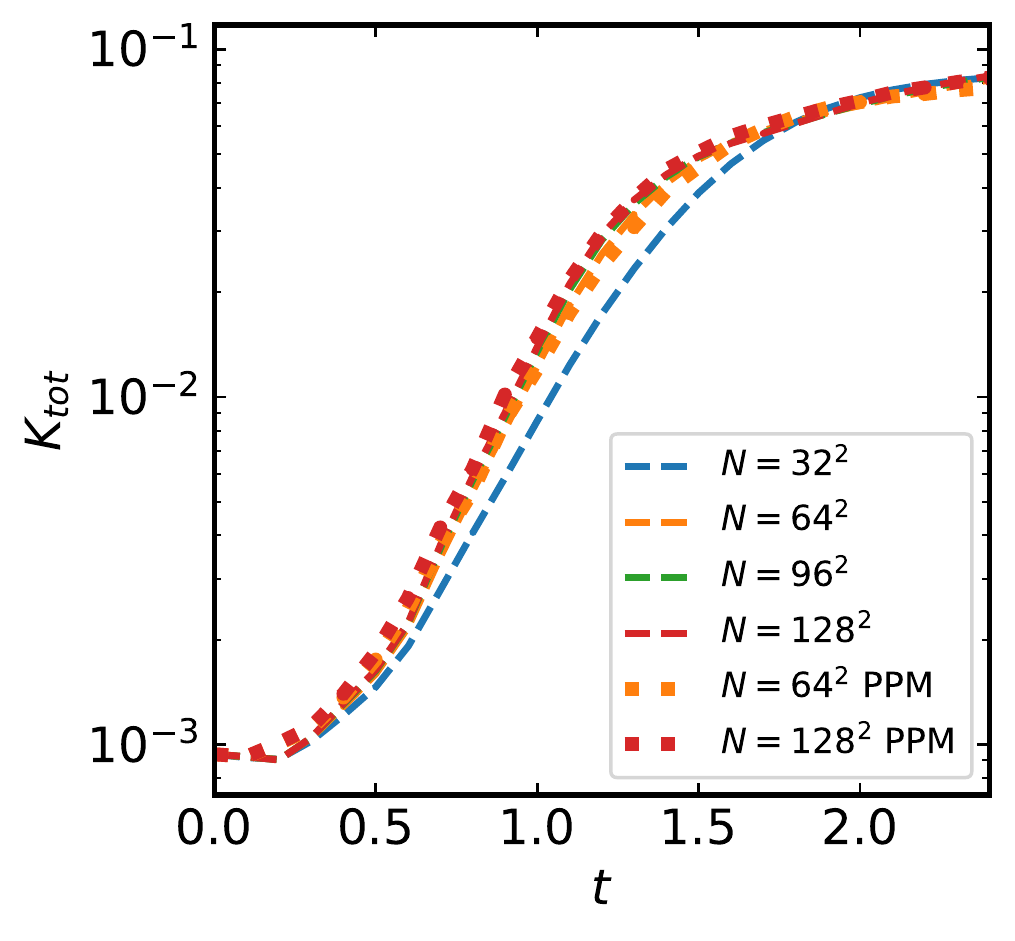}
    \caption{Total transverse kinetic energy for resolutions $N=32\time32$ (blue), $N=64\time64$ (orange), $N=96\time96$ (green), and $N=128\time128$ (red). LDA1 shown as dashed lines, PPM as dotted lines. The growth converges with resolution, and the non-linearity sets in earlier for higher resolution. The RD results are consistent with the well established PPM solver from Enzo.}
    \label{fig:tests_khy_ketot}
\end{figure}

The total kinetic transverse energy, found simply by summing the $y$-direction kinetic energy of each vertex, should grow exponentially \citep{ja:mcnally2012}. Figure \ref{fig:tests_khy_ketot}, which plots the total transverse kinetic energy $K_\mathrm{tot}$ for different resolutions, shows this growth between $t\approx0.3$ and $t\approx1.2$. The exponential growth appears linear in this log scale. Before this time, the kinetic energy is dominated by the initial sinusoidal perturbation. After this point, the growth becomes non-linear, as the initial turnover of the instability forms the more complex spiral structures. This time is earlier for higher resolutions, as finer structures are recovered.

Secondary instabilities can develop when using the sharp density discontinuity described above. These develop from the small variation in the boundary, created by the positions of the vertices. The variations create high frequency instability modes, which grow in higher resolution cases. In the low resolution cases, the same variations are present, but the higher numerical diffusion means that they dissipate into the background flow. In the test cases shown so far, the structured mesh has been used to make the boundary between flows as clean as possible. A random distribution of vertices would lead to a ragged edge, which has the potential to trigger instabilities at shorter wavelength modes, as discussed above, which make analysis of the results more difficult. A common technique to avoid this problem, in both random vertex cases, and high resolution uniform distribution setups, is to smooth the boundary layer with an exponential density and velocity profile \citep{ja:robertson2010, ja:lecoanet2016}. This ensures that the stimulated instability mode will dominate the evolution. The smoothed boundary is achieved by defining two new functions, $f(\theta)$ and $g(\theta)$. The first of these has the form
\begin{equation}
    f(\theta) = e^{-1/\theta},
\end{equation}
where theta is limited to $0\leq\theta\leq1$. The second function is given by
\begin{equation}
    g(\theta) = \frac{f(\theta)}{f(\theta) + f(1-\theta)}.
\end{equation}
Together these are used to smooth the density and velocity boundary between the sheer flows by setting the density and velocity respectively as
\begin{equation}
    \rho(y) = (\rho_0 - \rho_1)g\left(\frac{1}{2} + \frac{4y-1}{4d}\right)g\left(\frac{1}{2} - \frac{4y-3}{4d}\right) + \rho_1
\end{equation}
and
\begin{equation}
    v_x(y) = 2 v_{x0} g\left(\frac{1}{2} + \frac{4y - 1}{4d}\right) g\left(\frac{1}{2} - \frac{4y - 3}{4d}\right) - v_{x0}.
\end{equation}
The width of the boundary layer is dictated by $d$. All results shown from here on use this smoothed front, with a width of $d=0.1$.

\begin{figure}
    \centering
    \includegraphics[width=0.45\textwidth]{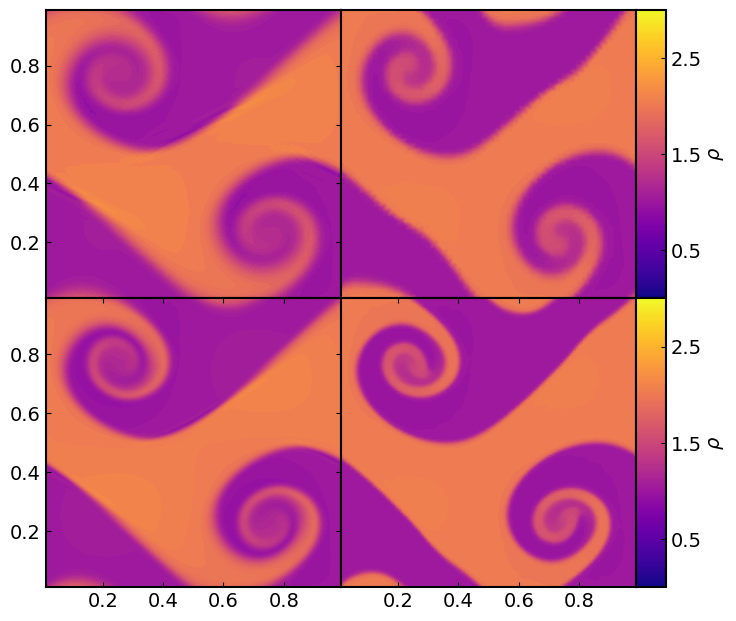}
    \caption{KH instability from the smooth initial conditions, comparing the LDA1 solver (left column), to the PPM solver (right column). Top row shows results for $N=64^2$, bottom row for $N=128^2$.}
    \label{fig:tests_khysmooth_comp}
\end{figure}

So far we have shown how well the different RD schemes perform with the KH test. In Figure \ref{fig:tests_khysmooth_comp} we show a comparison to the third order piece-wise parabolic method (PPM) used in Enzo \citep{ja:bryan2014}. The left hand column shows results from the LDA1 solver using $N=64^2$ and $N=128^2$ vertices. The right hand column shows results from PPM. These runs were produced using identical ICs. The PPM results are very similar to the our RD output. Both produce the expected spiral structure, with the PPM showing a well defined density contrast down to tighter turns of the spiral. This is thanks to a slightly lower numerical diffusion when using PPM. The total transverse kinetic energy, shown in Figure \ref{fig:tests_khy_ketot} with the orange ($N=64^2$) and red ($N=128^2$) dotted lines, grows equivalently to the RD LDA1 results, with only small differences, but following the same trend. As before, this transverse energy grows linearly in log space as expected, until the growth breaks down and becomes non-linear as the structure becomes more complex. The LDA1 growth shows good agreement with the well established Enzo results.

\begin{figure}
    \centering
    \includegraphics[width=0.45\textwidth]{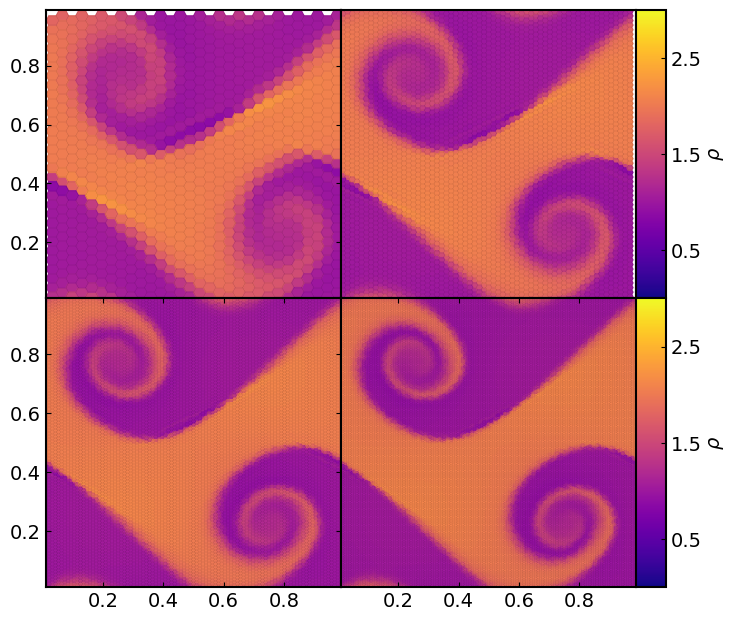}
    \caption{KH instability from smooth initial conditions, using LDA1, at different resolutions. These are $N=32^2$ (top left), $N=64^2$ (top right), $N=96^2$ (bottom left), and $N=128^2$ (bottom right). The higher mode instabilities are completely gone, and the stripe structures are significantly reduced.}
    \label{fig:tests_khysmooth}
\end{figure}

The RD solver's convergence with resolution is shown clearly in Figure \ref{fig:tests_khysmooth}, where I show panels for $N=32^2$, $N=64^2$, $N=96^2$, and $N=128^2$ vertices, again using the smoothed initial conditions. The results are consistent across the increasing resolutions, with the instability developing in the same place in each case. Even in the lowest resolution case, the spiral structure forms. The density contrast is present until the spiral is less than 3-5 vertices across, when the structure diffuses into the background. The higher resolution cases can resolve more detail within the instability, but even at low resolution the numerical solution retains a strong level of fine structure.

\subsubsection{Yee Isentropic Vortex}
Here we test the solver with the isentropic vortex. The density, velocity, and pressure profiles are chosen such that they are smooth, and that they produce an analytically stable solution \citep{ja:yee1999}. For a vortex centred on $(x_c,y_c)$, the velocity is given by
\begin{equation}
    v_x = -\Omega(r)(y-y_c) \hspace{5mm}\mathrm{and}\hspace{5mm} v_y = \Omega(r)(x-x_c)),
\end{equation}
where the radius $r$ is defined relative to the vortex centre, and $\Omega$ is the angular velocity profile
\begin{equation}
    \Omega(r) = \frac{\beta}{2\pi} \exp\left(\frac{1-r^2}{2}\right)
\end{equation}
with free parameter $\beta$. Taking pressure $p_\infty$ and density $\rho_\infty$ far from the vortex centre, the temperature profile is
\begin{equation}
    T(r)=\frac{p_\infty}{\rho_\infty}-\frac{\gamma-1}{\gamma}\frac{\beta^2}{8\pi^2}\exp \left(1-r^2\right).
\end{equation}
Based on this and normalisation constant $K$, the density profile becomes,
\begin{equation}
    \rho(r) = \left(\frac{T(r)}{K}\right)^\frac{1}{\gamma-1},
\end{equation}
and the pressure
\begin{equation}
    p(r) = K\rho(r)^\gamma.
\end{equation}
We take $\beta=5$, and $p_\infty=\rho_\infty=K=1$, use a uniform distribution of vertices, and run the setup to $t=10$. As the solution remains stable, the most effective comparison is made by looking at the L1 error function for the density, which, for a uniform vertex distribution, is given by
\begin{equation}
    L_1 = \frac{1}{N_\mathrm{side}}\sum_i |\rho_\mathrm{num,i} - \rho_\mathrm{ana,i}|,
\end{equation}
where $\rho_{\mathrm{num},i}$ is the numerical density associated with vertex $i$, $\rho_{\mathrm{ana},i}$ is the analytic solution, which in this case is simply the initial conditions, and $N_\mathrm{side}$ is the number of vertices per side.

\begin{figure}
    \centering
    \includegraphics[width=0.45\textwidth]{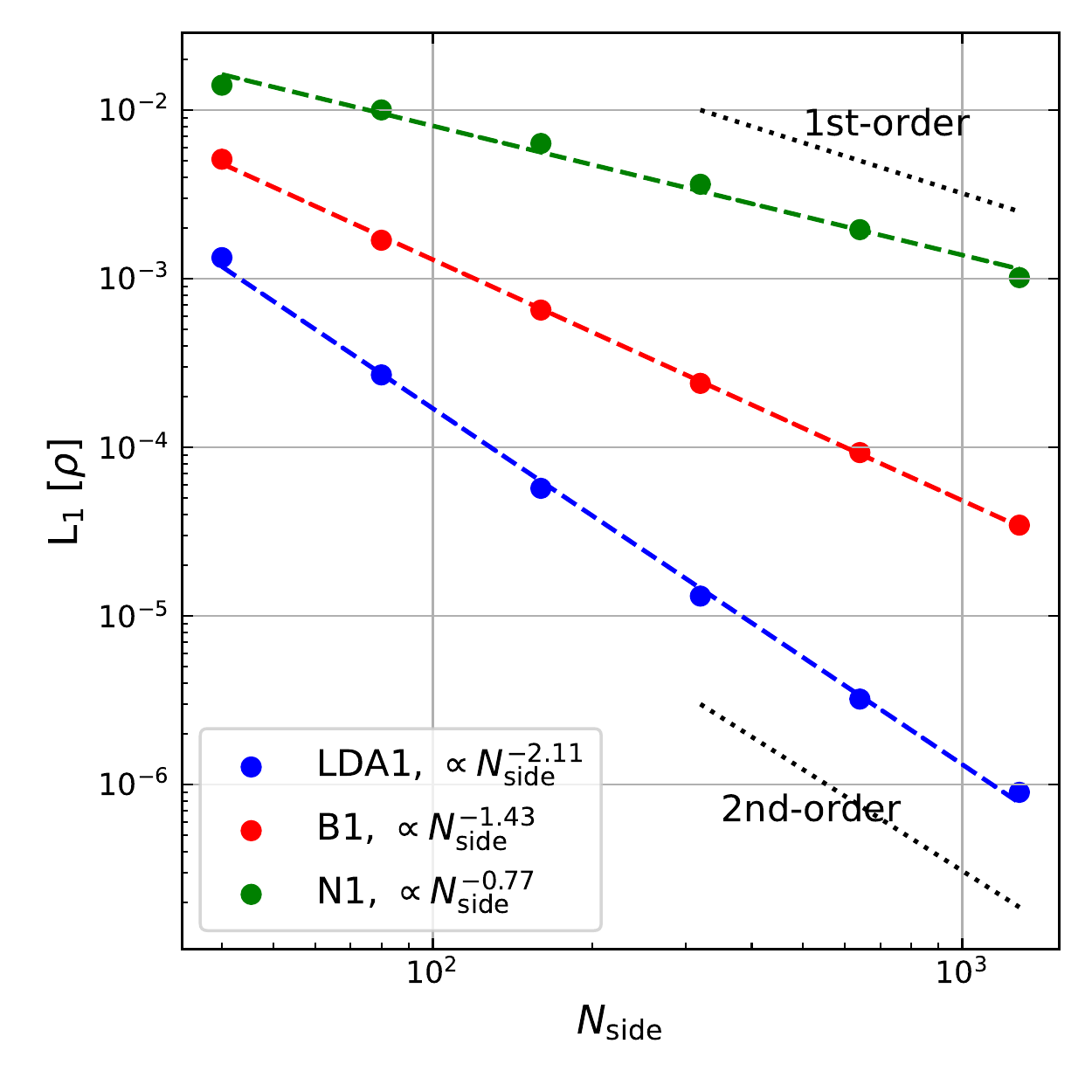}
    \caption{L1 error norm of density for the Yee isentropic vortex test, showing points for the LDA1 (blue), N1 (green), and B1 (red) schemes. Lines give the logarithmic fit, with the scaling relative to number of vertices per side $N_\mathrm{side}$ given in the legend. The LDA solver achieves second order accuracy in space.}
    \label{fig:yee_L1}
\end{figure}

We show the L1 error function results for the LDA1 (blue), N1 (green) and B1 (red) setups in Figure \ref{fig:yee_L1}. This figure also includes the fitted function for each distribution, using here $\mathrm{log}(L_1) = a \mathrm{log}(N_\mathrm{side}) + b$. The gradient of this fitted line in log space is the scaling of $L_1$ relative to the number of vertices per side $N_\mathrm{side}$. This scaling is given in the legend for each setup separately. This figure demonstrates that we can achieve second order accuracy in space with the LDA setup, but that this accuracy is dependent on the choice of distribution scheme. As expected, the N scheme is not second order accurate, due to Godunov's theorem. The blended scheme shows an order of accuracy half way between to two schemes that it blends together.

This test demonstrates that the RD solver can achieve second order accuracy in space, but only for certain choices of distribution scheme. With respect to the blended scheme used here, its order being between the two schemes that it blends suggests that it is not heavily favouring either. Since the problem is dominated by simple advection, with the structure remaining smooth throughout, it would be better if the LDA scheme were more heavily favoured. There is significant scope for implementing alternative blending mechanisms, such as by taking the minimum or maximum blending coefficient value of any point in the mesh \citep{ja:paardekooper2017}. Other blending schemes, such as the Bx scheme \citep{ja:dobes2008}, blend more aggressively, heavily weighting the N scheme in the presence of shocks by using a shock sensor based on the pressure gradient. This will be explored in future work.

\subsubsection{Sedov Blast}
The Sedov blast \citep{tb:sedov1959} replicates an explosion in a zero pressure environment. It reproduces conditions similar to the explosion from a highly idealised supernova. This is achieved with a static, uniform density and pressure, background medium at all positions. For this test case, we use a random distribution of vertices. The explosion is triggered by injecting a large amount of energy into the centre of the domain. In this case, we do this by setting the pressure to an extreme value. Ideally the pressure would only be injected into one vertex, to replicate a point explosion, but when this is done the initial propagation of the explosion can only follow the connections to the nearest vertices, leading to highly asymmetric wave. To avoid this, a circular region is defined, within which the energy is injected. The region is large enough that the outward flow is approximately radial, but small enough that the analytic solution is still applicable.

The explosion is expected to create a spherically expanding wave with a shock at the expansion front. The velocity at which the front move is set by the density of the background medium and the initial energy of the explosion, with the radius of the blast wave given by
\begin{equation}
    r(t) = \lambda\left(\frac{Et^2}{\rho_0}\right)^\frac{1}{5}.
\end{equation}
Here we denote the total energy of the explosion by $E$, and the background density with $\rho_0$. The coefficient $\lambda$ depends on adiabatic gas constant $\gamma$, at $\lambda\approx1.12$ for the $\gamma=5/3$ used here. Behind the shock front is an exponential density profile, falling to close to zero at the centre of the explosion.

\begin{figure}
    \centering
    \includegraphics[width=0.45\textwidth]{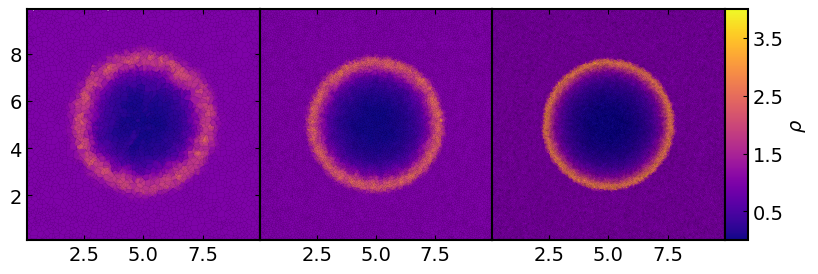}
    \caption{Sedov blast density results with increasing resolution, $N=64^2$ (left), $N=128^2$ (middle), and $N=256^2$ (right), all at $t=0.01$.}
    \label{fig:tests_sedov_res}
\end{figure}

\begin{figure}
    \centering
    \includegraphics[width=0.45\textwidth]{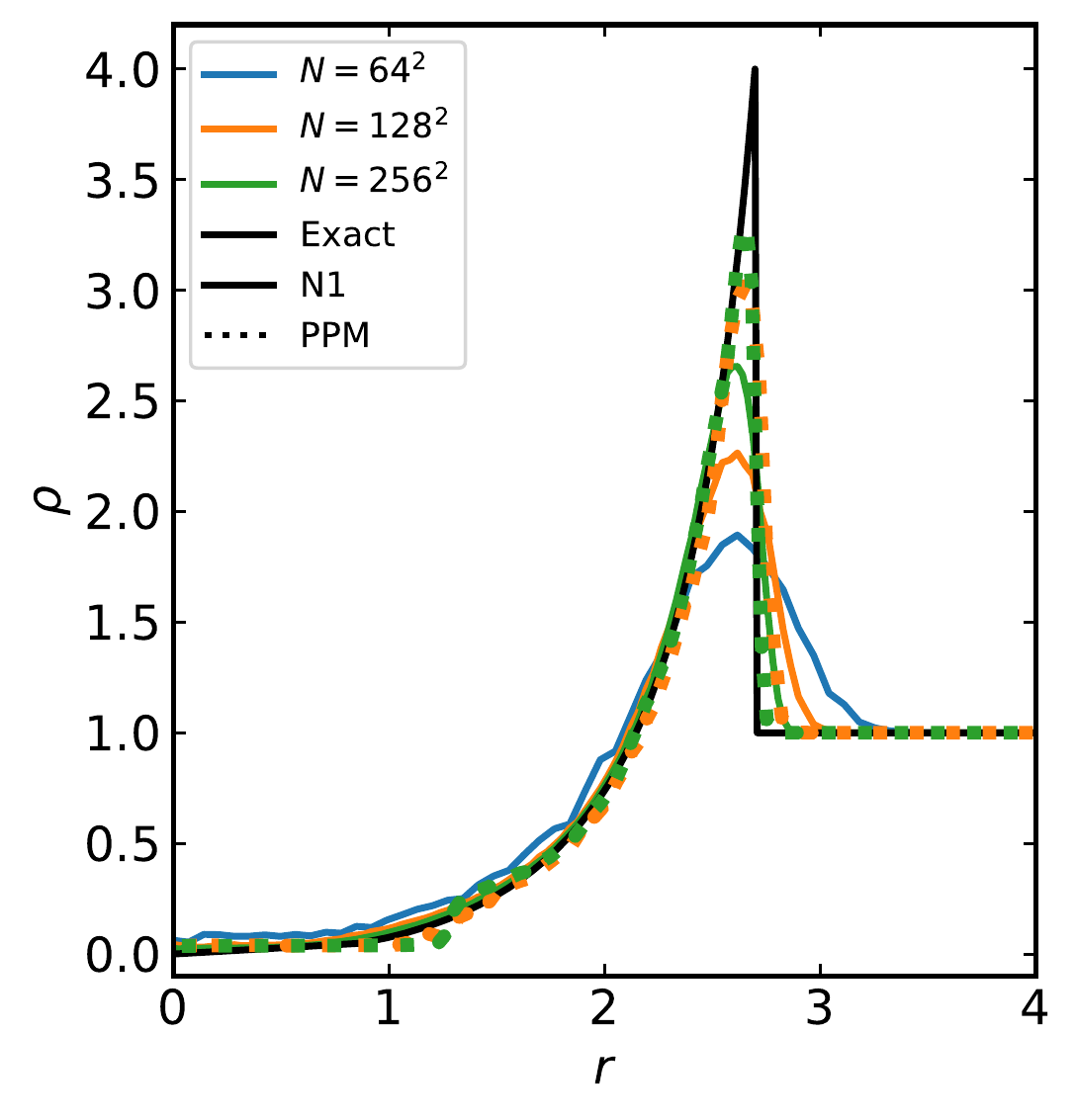}
    \caption{Radial density profiles, compared to the analytic prediction (solid black line). Results are shown for the three resolutions, $N=64^2$ (blue), $N=128^2$ (orange), and $N=256^2$ (green), with solid lines showing N1 results, and dotted lines showing PPM. The 3-5 resolution element limit, found in the KH tests, correspond to 0.4-0.8, 0.2-0.4, and 0.1-0.2 length units for these three resolutions.}
    \label{fig:tests_sedov_rad}
\end{figure}

In Figure \ref{fig:tests_sedov_res}, we show a comparison of the propagation of the explosion for three spatial resolutions, $N=64^2$ (left column), $N=128^2$ (middle column), and $N=256^2$ (right column), for the N1 solver. The results from the N2 scheme are effectively identical in this comparison, so are not shown. The LDA1 and LDA2 schemes do not produce stable results, due to their poor handling of strong discontinuities. The blended schemes heavily favour the N scheme in this test, and so produce results identical those shown for N1. The structure of the blast wave is recovered at all resolutions, with the dense wave sweeping up material as it moves radially outwards. As resolution is increased, the basic structure does not change, but the density profile does narrow. The narrower profile also shows a higher peak density. This is shown in more detail in Figure \ref{fig:tests_sedov_rad}. The key result from this comparison, however, is the consistency between resolutions. The blast wave is in essentially the same position at a given time, though the extent of the profile differs.

As before, we show a direct comparison to the PPM solver. Included in Figure \ref{fig:tests_sedov_rad}, as dotted lines, are profiles for the PPM solver, using $N=128^2$ (orange) and $N=256^2$ (green). These show sharper profiles at the explosion front, including higher peak values, consistent with the higher numerical diffusion present in the N scheme.

\subsubsection{Noh Problem}
The Noh problem \citep{ja:noh1987, ja:paardekooper2017} tests the ability of a solver to model the conversion of kinetic energy into internal energy. This is similar to the Sedov test, which features the conversion of internal energy to kinetic energy in the injection of energy through pressure. Once again, a random vertex distribution is used. It consists of a uniform density box, where the initial velocity at every position points radially inwards towards the centre of the box. The ideal initial conditions have zero pressure throughout the box. We use a cube box of side length $l=2$, with initial density $\rho_0=1$. When the pressure is exactly zero, the $K$-matrix becomes singular, and so cannot be inverted. Instead, we initialise the problem with a negligible, but non-zero, pressure of $P=10^{-6}$. The adiabatic gas constant is taken to be $\gamma=5/3$. With these initial conditions, the cylindrically symmetric exact solution \citep{ja:paardekooper2017} is as follows. At time $t$ after the start of the problem, the density at radius $r$ from the centre of the box is
\begin{equation}
    \rho(r,t) = 
    \begin{cases} 
        16 & \mbox{if } r < t/3 \\ 
        1 + t/r & \mbox{if } r \ge t/3
    \end{cases},
\end{equation}
the velocity magnitude is
\begin{equation}
    |v(r,t)| = 
    \begin{cases} 
        0 & \mbox{if } r < t/3 \\ 
        1 & \mbox{if } r \ge t/3
    \end{cases},
\end{equation}
and the pressure is
\begin{equation}
    P(r,t) = 
    \begin{cases} 
        16/3 & \mbox{if } r < t/3 \\ 
        0 & \mbox{if } r \ge t/3
    \end{cases}.
\end{equation}
These equations describe the build up of a uniform density cylinder in pressure equilibrium. The cylinder expands as material flows towards the centre, with a shock at its surface. Material outside the cylinder continues to flow inwards at its initial rate.

\begin{figure}
    \centering
    \includegraphics[width=0.45\textwidth]{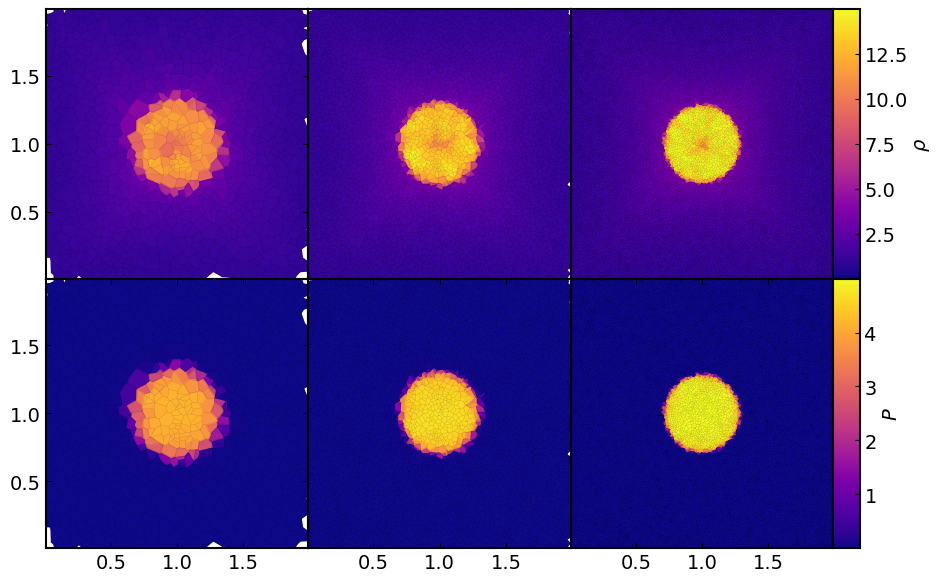}
    \caption{Density (top row) and pressure (bottom row) distributions from the Noh problem at $t=0.8s$, using the N1 solver, for resolutions using $N=32^2$ (left), $N=64^2$ (middle), and $N=128^2$ (right) vertices.}
    \label{fig:tests_noh}
\end{figure}

We show the difference in the density and pressure results with increasing resolution, for the first order N scheme, in Figure \ref{fig:tests_noh}, with $N=32^2$, $N=64^2$, $N=128^2$ vertices, at $t=0.8$. For this test, we have chosen to also look at the solvers abilities at very low resolution, hence including the $N=32^2$ case. We only show the N1 solver here, because the LDA solver struggles with the extreme conditions at the very centre of the box, when the shock first forms. As before, the blending scheme strongly favour the N scheme in this scenario. The N2 results are not significantly different for this test. The blending scheme used here does not favour the N scheme strongly enough to counteract the LDA schemes struggle with this test. A difference blending with a stronger weighting mechanism, such as the Bx scheme may perform better. This will be explored in future work.

The density inside the shock increases with resolution, suggesting this density is somehow dependent on the formation of the shock at the centre of the box. At lower resolution, this initial radius will be larger, since the elements are larger and the central shock will form over a wider physical area. The shock front at the surface sharpens with the increasing resolution. The shock is resolved by  approximately 3-5 mesh vertices, represented here by their dual cells. There is some variation in density within the cylinder, particularly in the centre, where there is a small low density cavity. This region gets smaller with increased resolution. This is likely an artifact of the finite resolution. When the initial flow builds up material in the innermost region, material can only flow in a small number of directions, limited by the exact structure of the mesh. The inward radial flow is therefore not well resolved, leading to this less exact solution. The position of the shock is not well defined when the circular shape is only resolved by a few vertices. The pressure, on the other hand, is significantly more uniform across the whole cylinder, demonstrating the expected pressure equilibrium. The lower density and equal pressure show that the temperature in the inner region must be higher. This heating phenomenon is known as `wall heating', and has been previously identified in a number of Riemann-type hydro-solvers \citep{ja:noh1987, ja:rider2000, ja:stone2008}.

\begin{figure}
    \centering
    \includegraphics[width=0.4\textwidth]{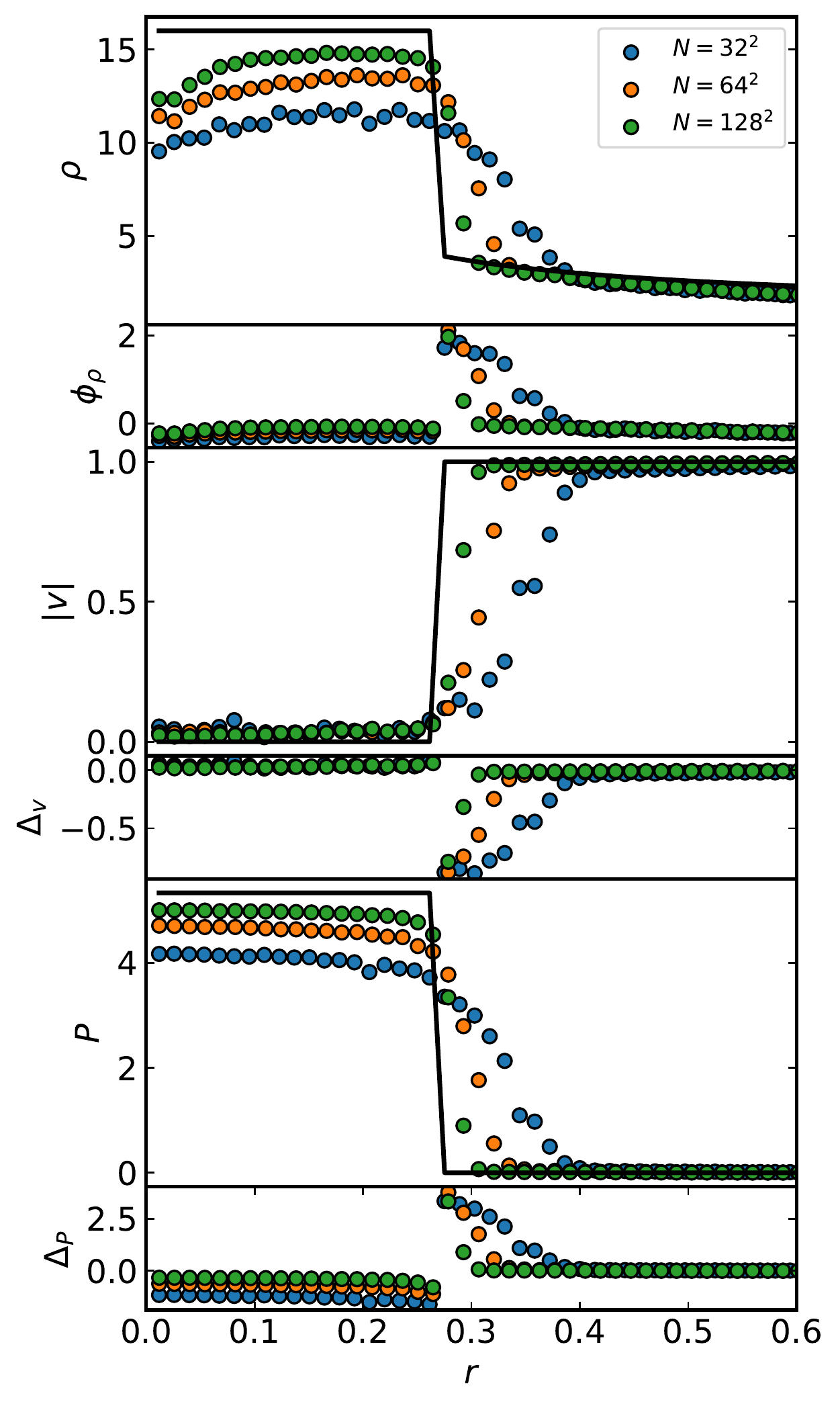}
    \caption{Radial profiles from the Noh problem, showing the density (top), velocity magnitude (middle), and pressure (bottom). The blue dots show the results for $N=32^2$, the orange show $N=64^2$, the green show $N=128^2$, and the red $N=256^2$. The black line shows the exact solution. Below each profile is the either the residual $\phi = (X_\mathrm{num} - X_\mathrm{exe})/X_\mathrm{exe}$, or the difference $\Delta = (X_\mathrm{num} - X_\mathrm{exe})$, if the analytic solution includes significant regions with zeros.}
    \label{fig:tests_noh_profile}
\end{figure}

\subsubsection{Gresho Vortex}
\begin{figure}
    \centering
    \includegraphics[width=0.4\textwidth]{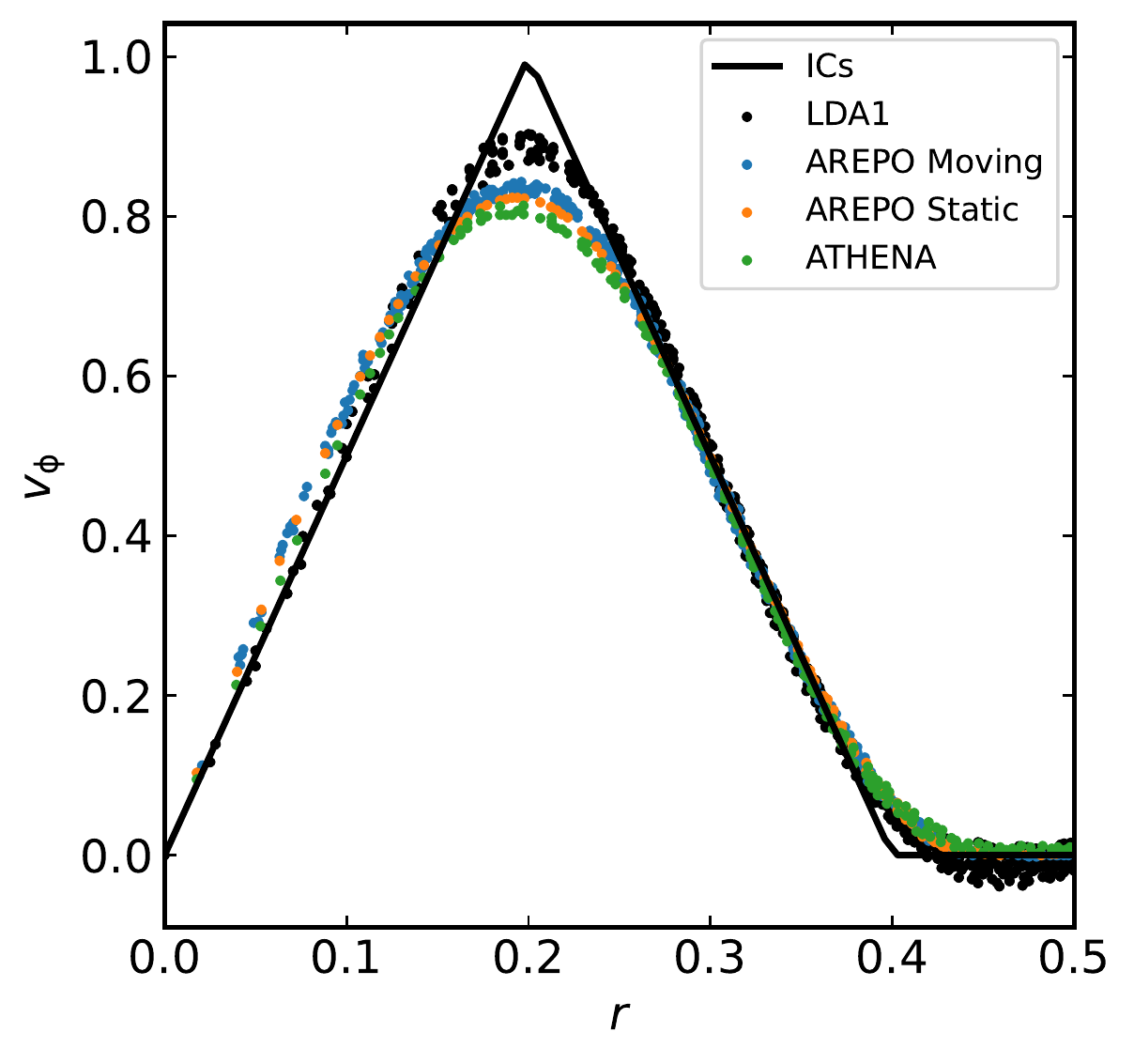}
    \caption{Radial profile of the azimuthal velocity at $t=3$. The black points show the LDA1 output, using $N=40^2$ vertices, with the initial conditions shown as the solid line. The blue (AREPO moving mesh), orange (AREPO static mesh), and green (ATHENA) points show results taken from Figure 28 of \citet{ja:springel2009}. The LDA1 results maintain a peak value closer to the initial conditions.}
    \label{fig:tests_gresho_profile}
\end{figure}

We can test the conservation of angular momentum within the solver by considering the Gresho vortex problem \citep{ja:gresho1990, ja:liska2003}, which consists of a stable rotating region, embedded in a uniform density box. The region is defined by setting up a rotational velocity profile $v_\phi$,which varies with radius $r$, such that
\begin{equation}
    v_\phi(r) = \left\{
    \begin{array}{ccc}
        5r & \mathrm{for} & 0 \leq r < 0.2  \\
        2-5r & \mathrm{for} & 0.2 \leq r < 0.4 \\
        0 & \mathrm{for} & r \geq 0.4
    \end{array}\right.,
\end{equation}
with a uniform background density of $\rho=1$. This rotation is stable for an appropriate pressure profile \citep{ja:liska2003}
\begin{equation}
    P_\phi(r) = \left\{
    \begin{array}{ccc}
        5+25/2r^2 & \mathrm{for} & 0 \leq r < 0.2  \\
        9+25/2r^2-20r+4\ln(r/0.2) & \mathrm{for} & 0.2 \leq r < 0.4 \\
        3+4\ln2 & \mathrm{for} & r \geq 0.4
    \end{array}\right..
\end{equation}
However, the finite resolution of any numerical scheme will introduce small inaccuracies in this pressure, leading to a potential loss of stability. In Figure \ref{fig:tests_gresho_profile}, we show the azimuthal velocity at $t=3$, for the LDA1 scheme, using $N=40^2$ vertices. Here we use a uniform distribution of vertices. The black points represent the numerical output, with the initial conditions given by the solid line. The RD solver performs well, maintaining much of the sharp profile. In particular, we note the height of the peak in the rotational velocity, at approximately $v_\phi = 0.9$. The closer this is to the initial value of unity, the better. The smoothing of the sharp peak in velocity is a manifestation of the numerical diffusion discussed before.

Our results are competitive with those published for this test from other modern astrophysical codes, such as the results shown for AREPO \citep{ja:springel2009}, in both static and moving mesh modes, for their zero bulk motion case. Our results are also competitive with, ATHENA \citep{ja:stone2008}, the code compared to in the AREPO paper, with ATHENA results shown for a 2D Roe solver implementation. It is interesting to compare the performance against our own directly therefore, due to the developmental link between the methods. Taking the peak of the ATHENA Gresho velocity profile as $v_\phi=0.8$ \citep[see top right panel of Figure 28 of][]{ja:springel2009}, we can directly compare the truly multi-dimensional RD solver to a close dimensional-splitting equivalent. This comparison is shown explicitly in Figure \ref{fig:tests_gresho_profile}, where the results for the Roe solver in ATHENA, and the moving mesh and static modes of AREPO, are shown in green, blue and orange respectively. LDA1 performs well in this test when compared to all three alternatives. The RD results show as little as half the level of smoothing, possibly  a consequence of the truly multi-dimensional nature of the approach.

\subsection{3D Tests}
A key feature of this implementation is extension to full 3D, something not found in current RD solvers used in astrophysics \citep{ja:paardekooper2017}. The truly multidimensional nature of the solver is of even greater importance, in 3D, than in the 2D case. Here we show results found using the 3D mode of the solver.

\subsubsection{Blob Test}
The blob test combines both Kevin-Helmholtz instabilities and Rayleigh-Taylor instabilities by embedding a cold cloud within a hot flow \citep{ja:agertz2007}. This test consists of a high density, static, spherical cloud $\rho_H$, placed within a low density background $\rho_L$ which moves with a bulk velocity $v_0$. In 2D this cloud is represented by a disk. The high density region is an order of magnitude more dense than the background medium. The whole region is in pressure equilibrium, with the low density wind much hotter than the cold cloud. The background medium is given a supersonic initial velocity, with Mach number $\mathcal{M}=v_0/c_s$, where $c_s$ is the sound speed of the gas. Astrophysically, this corresponds to high density clouds moving with relatively supersonic velocity through a lower density background, such as a region of cold ISM close to a supernova. The vertices are distributed randomly.

The initial linear stages of the evolution of this set up can be predicted with some degree of confidence \citep{ja:agertz2007}. The collision of the supersonic flow with the static density front will produce a bow shock upwind of the cloud, with a subsonic region behind the front. The cloud itself will be accelerated by its interaction with the flow. Kelvin-Helmholtz (KH) instabilities, discussed in isolation in Section \ref{sec:tests_2D}, build at the boundaries between shear flows, such as the boundary between the cloud and the background medium, where the radial vector is orthogonal to the flow. At the same time, Rayleigh-Taylor (RT) instabilities evolve where the cloud is pushed into the downwind low density medium \citep{tb:chandrasekhar1961}. Together these instabilities lead to the breakup of the original cloud. Tendrils of high density are pushed downwind, and the original sphere is crushed by the incoming flow, and is eventually destroyed. The lower limit of the time for this to happen is predicted by the crushing time \citep{ja:agertz2007}
\begin{equation}
    t_\mathrm{cr} = \frac{2r_\mathrm{cl} \chi^{1/2}}{v_0},
\end{equation}
where $r$ is the radius of the cloud, $\chi$ is the initial density contrast, and $v$ is the relative velocity of the cloud and the background flow. The crushing time comes from the time it takes for the wind to cross the extent of the cloud, scaled by the ratio of cloud density to wind density. A greater difference will result in a longer time to disrupt. This can be used as a reasonable gauge of time scale for the cloud to be disrupted. The full physical evolution of the cloud is highly non-linear, and so cannot be easily predicted.

We choose values of $\rho_\mathrm{cl}=100 \mathrm{kg/m}^3$, for the cloud, and $\rho_0=100 \mathrm{kg/m}^3$ for the background. The cloud extend $r=1\mathrm{m}$ from its centre. The box dimensions are $10\mathrm{m}\times10\mathrm{m}\times10\mathrm{m}$, and the initial velocity has Mach number $\mathcal{M}=1.5$, which corresponds to $v_0=6.2\mathrm{m/s}$. We run a number of such setups, varying resolution and solver type. We compare the predicted evolution to the results from these runs.

\begin{figure}
    \centering
    \includegraphics[width=0.45\textwidth]{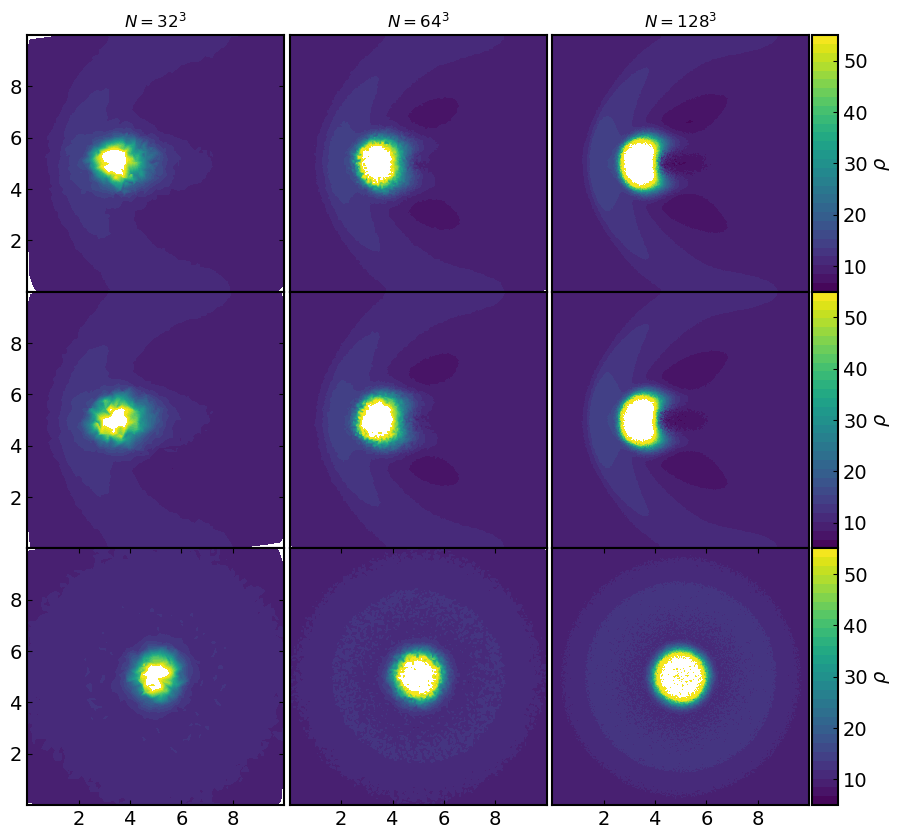}
    \caption{Blob test results for the N1 solver, with randomly distributed vertices, from the $N=32^3$ (left column), $N=64^3$ (middle column), and $N=128^3$ (right column) solver, at $t=t_\mathrm{cr}$. Each row shows the slice through the centre of the blob, with the X-Y plane in the top row, the X-Z plane in the middle row, and Y-Z in the bottom row.}
    \label{fig:tests_3D_blob}
\end{figure}

In Figure \ref{fig:tests_3D_blob}, we show results for the density distribution, with using the N1 solver, with $N=32^3$ (left column), $N=64^3$ (middle column), and $N=128^3$ (right column) vertices, at $t=t_\mathrm{cr}$. As with the Sedov results, each row shows a different plane, through the centre of the cloud. From top to bottom, these are X-Y, X-Z and Y-Z. Unlike the Sedov case, the evolution is not spherically symmetric. The bottom row of panels show the head on view of the cloud, showing the symmetry in the other dimensions. Even at  low resolution, we see the development of the bow wave, and some disruption of the cloud itself. The edges of the cloud being shredded by instabilities, and material from the cloud is accelerated by the wind, as expected. We only show N1 results here, as the second order and blended results are essentially identical.

As we increase the resolution first to $N=64^3$, and then to $N=128^3$, we see a more finely structured bow shock. The head on view (bottom row) shows this clearly. The RT instabilities behind the cloud are more clearly seen here as well, with the greater number of resolution elements recovering the effect in more detail. The higher resolution is also able to recover the low density regions behind the bow wave with much greater detail. These resolutions were chosen to test how well the solver can handle such cloud break ups at very low resolution, as in many of the eventual simulation scenarios, the analogous clouds are only resolved at similar or lower resolutions.

\begin{figure}
    \centering
    \includegraphics[width=0.45\textwidth]{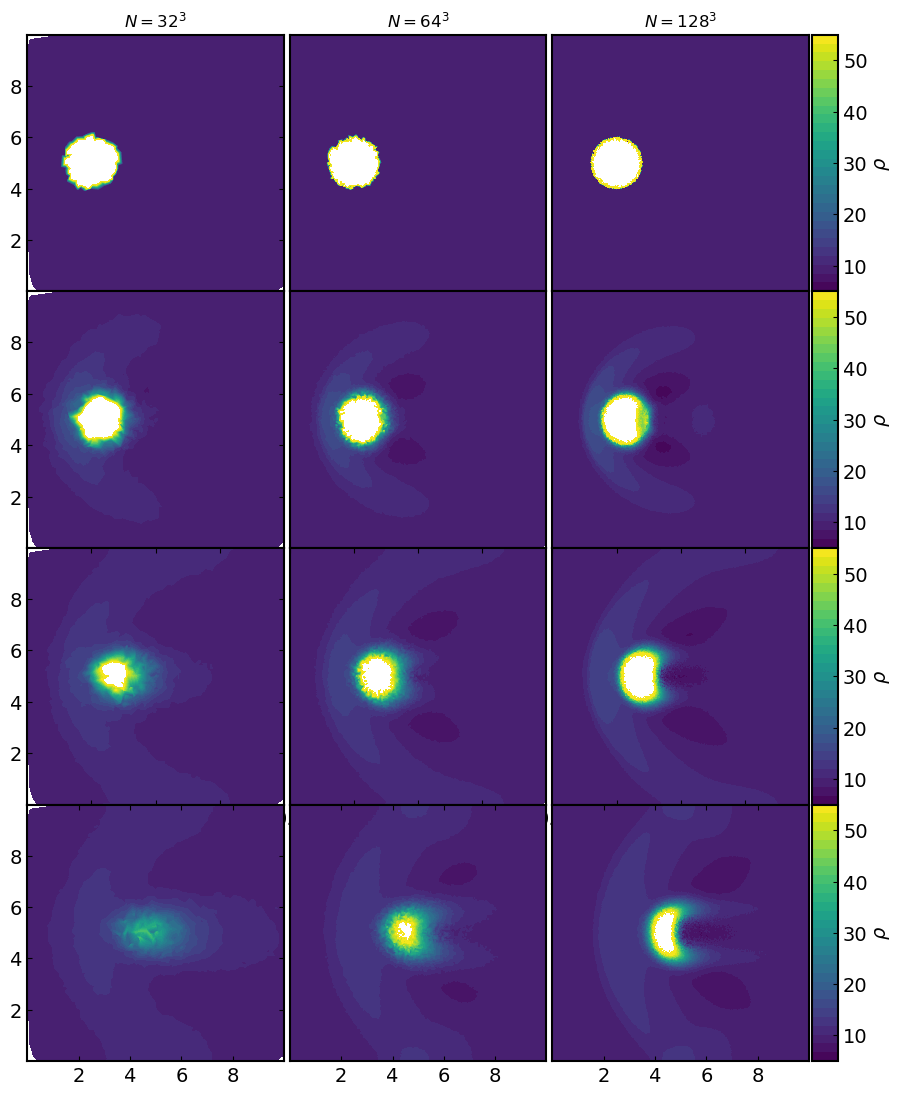}
    \caption{Evolution of the cold gas cloud using the N1 solver, for $N=32^3$ (left column), $N=64^3$ (middle column), $N=128^3$ (right column) vertices. From top to bottom, each row is taken at time $t=0$, $t=0.5t_\mathrm{cr}$, $t=t_\mathrm{cr}$, and $t=1.5t_\mathrm{cr}$ respectively. The bow wave builds as the hot flow collides with the cloud, with wings extending to the edge of the box. The extent of the disruption depends on the resolution, with the low resolution case struggling to resist breakup and diffusion into the surroundings, while the higher resolution cases survive longer.}
    \label{fig:tests_3D_blob_time}
\end{figure}

We also show the time evolution of the results. In Figure \ref{fig:tests_3D_blob_time} we compare the evolution of the density distribution for $N=32^3$ (left column), $N=64^3$ (middle column), and $N=128^3$ (right column) blob tests. Time increases downwards, with the first row showing the initial conditions at $t=0$, the second at $t=0.5t_\mathrm{cr}$, the third at $t=t_\mathrm{cr}$, and the fourth at $t=1.5t_\mathrm{cr}$. The progressive build up of the bow wave is clear, as is the acceleration of the cloud by the hot flow. The low resolution case shows a greater amount of mixing, between the cloud and the background, by $t=1.5t_\mathrm{cr}$, while the higher resolution cloud retains some of its integrity. This is likely a direct effect of the small number of resolution elements that make up the $N=32^3$ cloud. Once again, it is clear we have not converged with resolution. The fundamental evolution is present across all cases, showing the solver performs well with these highly multi-dimensional flows, even when only given a small number of vertices with which to work.

\begin{figure}
    \centering
    \includegraphics[width=0.45\textwidth]{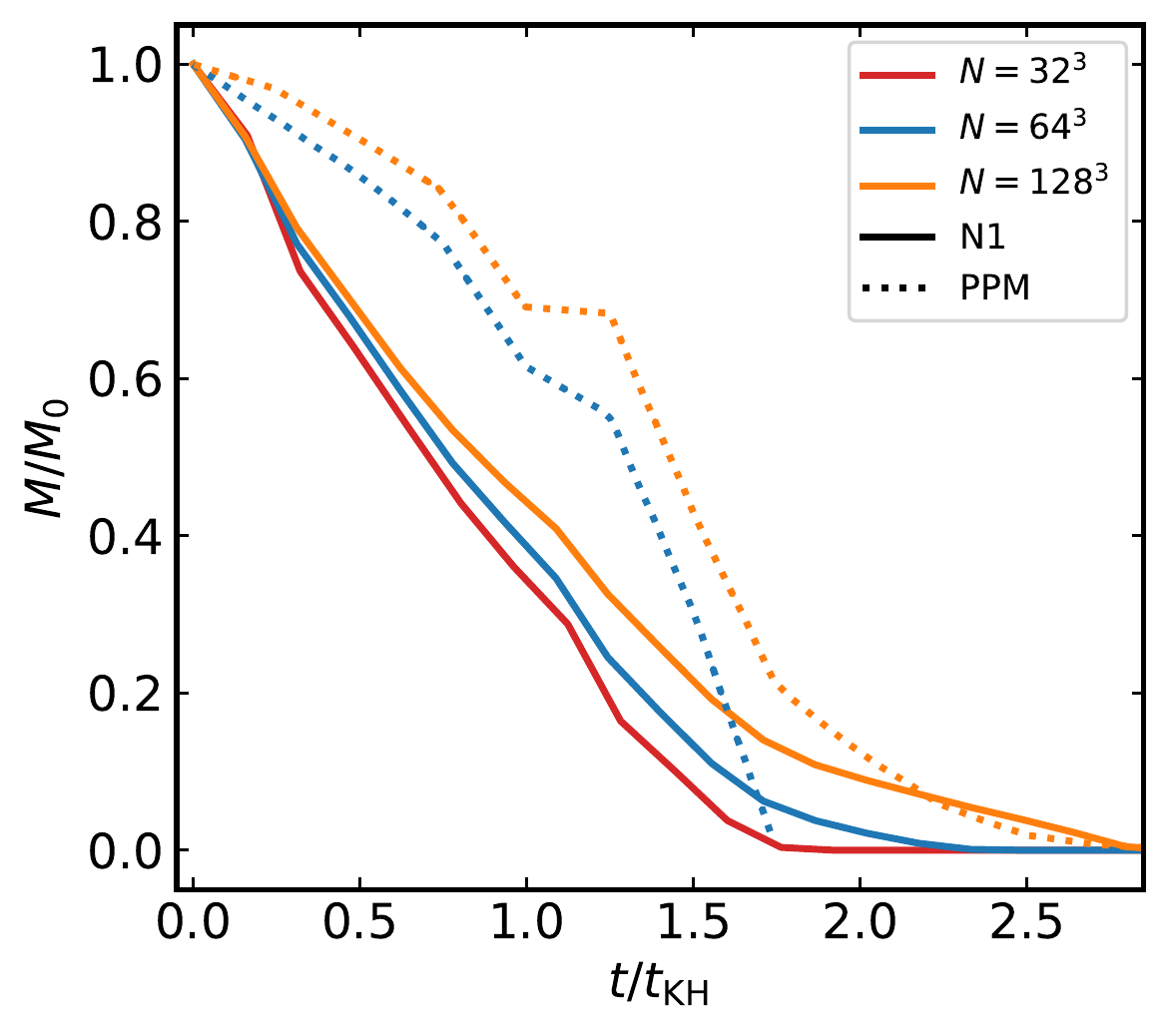}
    \caption{Evolution of cloud mass using $N=32^3$ (red), $N=64^3$ (blue), and $N=128^3$ (orange), comparing the N1 (solid) solver to the PPM solver (dotted). N1 clouds disperse more rapidly than those modelled using the PPM solver, largely independent of resolution, but the final destruction time is in better agreement.}
    \label{fig:tests_3D_blob_cloud_mass}
\end{figure}

The similarities, and differences, are emphasised when we look at the mass of gas in the cloud. Figure \ref{fig:tests_3D_blob_cloud_mass} compares the evolution of the cloud mass, in the RD solver results, to that from the PPM solver \citep[taken from][]{ja:agertz2007}, and at different resolutions. From here we normalise by the characteristic growth time of the Kelvin-Helmholtz instabilities $t_\mathrm{KH}=1.6t_\mathrm{cr}$, to allow direct comparison between the works. We follow the cloud definition in \citet{ja:agertz2007}, where mass is assigned to the cloud if the density at the element position is $\rho>0.64\rho_\mathrm{cl}$, and the temperature is $T<0.9T_\mathrm{ext}$, where $T_\mathrm{ext}$ is the initial temperature of the background medium. The cloud mass, in N1 case (solid lines), diminishes smoothly to zero, with the shortest complete depletion time, in the lowest resolution case, coming in at $t\sim1.6t_\mathrm{KH}$, while the highest resolution case survives the longest, only reaching zero mass at $t\sim2.8t_\mathrm{KH}$. If we take the somewhat arbitrary, but useful, point of cloud disruption as being when it has lost half its mass, this spread is significantly reduced, ranging from $t\sim0.70t_\mathrm{KH}$ to $t\sim0.86t_\mathrm{KH}$. This resolution dependence is consistent with previous results \citep{ja:agertz2007}, such as those shown for the Enzo PPM solver (dotted lines). However, the PPM clouds present different evolutionary paths, when compared to our RD results. The 0.5 depletion times for $N=64^3$ and $N=128^3$ cases are, respectively, $t\sim1.3t_\mathrm{KH}$ and $t\sim1.4t_\mathrm{KH}$. It is not completely clear why the RD cases have clouds that are disrupted so much more quickly, or why we do not see the same detailed evolution of the cloud structure. The different solvers agree more closely on the final destruction time, with the PPM $N=64^3$ and $N=128^3$ case reaching zero at $t\sim1.7t_\mathrm{KH}$ and $t\sim2.8t_\mathrm{KH}$ respectively. The $N=128^3$ case in particular show a very similar final destruction.

The RD results do not show the same extent of lateral elongation, instead largely retaining their spherical shape, except in the $N=128^3$ case, where there is some evidence of this elongation, but only at $t=1.5t_\mathrm{cr}$. We also do not see the fragmentation of the detached material. Instead material diffuses into a smooth wake at all resolutions. This can likely be explained again by the high numerical dissipation of the N scheme (and the blended scheme for this problem, as it heavily favours the N scheme). In this context, numerical dissipation causes the cloud to spread material as it is forced back by the momentum transferred from the background medium. This is partially borne out by the extended lifetime of at the higher resolutions, but the increase is small. A more detailed study on the cause for these differences will be presented as part of a future study focused on blending schemes within the RD method.

This test is fundamentally different from the Sedov blast, in that it is truly three dimensional. While the Sedov blast has flows in all three cardinal directions, it remains spherically symmetric. Therefore the 3D test is not significantly different from its 2D counterpart. The blob test, on the other hand, contains both KH and RT instabilities. A given mode of the KH instability can characterised by only one wave number, and as the vortices form between the shear flows, they rotate on a plane. In 2D, there is only one plane in which they can form, but in 3D there is an additional degree of freedom, complicating the growth of this instability. The RT instabilities can form in either 2D or 3D, requiring up to two wave numbers to characterise a given mode \citep{ja:agertz2007}. The effect of these differences is that the disruption of the blob will proceed in an inherently different manner in 3D, as opposed to 2D. It is also a structurally complex evolution that is a strong test the overall capabilities of a solver.

\section{Discussion and Conclusions}

In this paper, we introduce the fundamental concepts behind the development of the residual distribution hydrodynamics solvers, including the one dimensional equivalent, the Roe solver, and the 2D and 3D forms of the hydro solver. This solver is truly multi-dimensional, as it requires no dimensional splitting, and contains a whole family of methods that are all built around the same base: calculating a residual over a triangular element, in a single calculation, and then distributing this to the vertices of the element, to update the solution to the set of PDEs that are being solved. We cover the various choices that can be made when designing a specific implementation of such a solver, which define the resultant characteristics and abilities of the code. This includes the required linearisation of the Euler fluid equations, first laid out by Roe. We also introduce the Delaunay triangulation, with a brief description of its definition, properties and construction. We discuss the extensive testing we performed on the RD solver implementation, covering one, two, and three dimensional test cases. These tests demonstrate the strengths of the solver in recovering multi-dimensional flows, while also handling shocks and other extreme situations well. The RD implementation that we describe, and test, here, represents what the solver can do without significant optimisation, or tailoring to a given problem. It performs well when compared to current stat-of-art solvers, and can resolve complex structures at low resolution. In particular, it demonstrates a strong ability to maintain stability in the Gresho test.

There is still significant scope to optimise the RD solvers to any desired problem, with a straightforward framework for implementing different distribution and blending schemes. The fact that it is built around an unstructured mesh makes it the perfect candidate for conversion into a moving-mesh scheme. It thus has great potential for further improvement. While beyond the scope of this paper, we do plan to extend our implementation to an arbitrary Lagrangian Eulerian (ALE) form \citep{ja:arpaia2015}, where the underlying mesh moves approximately with the physical flow of the fluid. Our expectation is that this will significantly reduce numerical dissipation, and our preliminary work confirms this. We will publish this ongoing work in a future paper. Overall, the RD solvers presented in this implementation are well on their way to being powerful new tools for running astrophysical simulations of a range of scenarios.

\section*{Acknowledgements}

The authors would like to thank Sijme-Jan Paardekooper for our useful discussions on the RD approach and possible time-stepping mechanisms, and Claudio Dalla Vecchia, and Isaac Alonso Asensio for their contributions to testing and comparing our results to their own. Ben Morton \& Sadegh Khochfar are grateful for support from the UK STFC via grants ST/V000594/1 and 564000 J56415. Zhenyu Wu thanks the Edinburgh-CSC scholarship for their support.

For the purpose of open access, the author has applied a Creative Commons Attribution (CC BY) licence to any Author Accepted Manuscript version arising from this submission.

\section*{Data availability}

The data underlying this article will be shared on reasonable request to the corresponding author.




\bibliographystyle{mnras}
\bibliography{master} 



\appendix

\section{Summary of RD Derivation} \label{app:RD_derivation}
Here we summarise the derivation of the 2D form of the RD solver, first for the first order form, and then for the second order. We also cover the detail of how the Euler equations are transformed for use with an RD solver. 

\subsection{Residual distribution in 2D - 1st Order} \label{app:RD_derivation_1st}
Starting from Equation (\ref{eq:method_2D_jac}), the spatial differential part of the PDE can be written in its discretised form as
\begin{equation}
    \frac{\partial \mathbf{Q}}{\partial x} = \frac{1}{2|T|}\left(\sum_{i=1}^3 \mathbf{Q}_i\mathbf{n}_i \right) \cdot \hat{\mathbf{x}},
\end{equation}
where the sum is over the three vertices of the element, $\mathbf{Q}_i$ is the state at each vertex, and $\mathbf{n}_i$ is the inward pointing normal to the edge opposite the vertex $i$. The unit vector in the $x$-direction is $\hat{\mathbf{x}}$. This amounts to finding the linear interpolation of the state in the element from the state at the three vertices. These partial differentials are now constant across the element. The Jacobian is a little harder to define. Substituting in the above equation, the residual becomes
\begin{equation}
    \begin{split}
        \phi^T = \frac{1}{2|T|} \left[\left(\sum_{i=1}^3 \mathbf{Q}_i\mathbf{n}_i \right) \cdot \hat{\mathbf{x}} \int_T \mathbf{A}_x  dxdy \right. \\ + \left. \left(\sum_{i=1}^3 \mathbf{Q}_i\mathbf{n}_i \right) \cdot \hat{\mathbf{y}} \int_T \mathbf{A}_y  dxdy \right].
    \end{split}
\end{equation}
Analogous to the Roe solver, it is possible to define an average Jacobian. In the 1D case, this was the average at the boundary, and in this 2D case it is the average over the element. This is therefore defined as the integral of $\mathbf{A}_x$ and $\mathbf{A}_y$ over the area of the element, divided by that area
\begin{equation}
    \bar{\mathbf{A}}_x = \frac{1}{|T|} \int_T \mathbf{A}_x dxdy,
\end{equation}
which leaves the residual as
\begin{equation}
    \phi^T = \frac{1}{2} \left[\left(\sum_{i=1}^3 \mathbf{Q}_i\mathbf{n}_i \right) \cdot \hat{\mathbf{x}}  \bar{\mathbf{A}}_x + \left(\sum_{i=1}^3 \mathbf{Q}_i\mathbf{n}_i \right) \cdot \hat{\mathbf{y}} \bar{\mathbf{A}}_y  \right].
\end{equation}
The dot product of the normal with the unit vectors mean only that component is included in each sum. Using this, combining the summations, and bringing the Jacobian inside the sum gives
\begin{equation} \label{eq:method_2D_res_disc}
    \phi^T = \frac{1}{2} \sum_{i=1}^3 \left(\mathbf{Q}_i \bar{\mathbf{A}}_x n_{x,i} +  \mathbf{Q}_i\bar{\mathbf{A}}_y n_{y,i} \right).
\end{equation}

Finally, it is possible to combine the Jacobians into a single term $\mathbf{K}_i = (\bar{\mathbf{A}}_x n_{x,i}+ \bar{\mathbf{A}}_y n_{y,i})/2$, with the dependence on vertex $i$ coming from the normal of the opposite edge. This simplifies the calculation of the element residual to the sum of the product of matrix $\mathbf{K}_i$ and state $\mathbf{Q}_i$
\begin{equation}
    \phi^T = \sum_{i=1}^3 \mathbf{K}_i \mathbf{Q}_i.
\end{equation}
Now all that remains is a method to calculate the discrete form of the element Jacobian $\bar{\mathbf{A}}$. Since it has been required that the system of equations is linear, then the Jacobian $\mathbf{A}_x = \partial \mathbf{F}/\partial \mathbf{Q}$ will vary linearly. This means the element Jacobian can simply be computed as the Jacobian as a function of the average state of vertices of that element $\bar{\mathbf{A}}_x = \mathbf{A}_x \left(\bar{\mathbf{Q}}\right)$, where the average state is simply the mean of vertex states. 

The combining of the Jacobian into $\mathbf{K}_i$ is the key step in making this a truly multi-dimensional method. A similar method that considers the $x$ and $y$ Jacobians separately would effectively be splitting the problem by dimension.

\subsection{Residual distribution in 2D - 2nd Order} \label{app:RD_derivation_2nd}
The first order RD methods were largely developed to treat steady problems (i.e. ones where the solution converges on some steady state) \citep{ja:paillere1995, ja:hubbard1997, ja:dobes2008}. For these methods, having only first order accuracy in time is acceptable, but for problems with significant time variation, it is important to achieve second order accuracy in time. The order in time refers to the highest order of term included in the approximation of the solution. The solution can be written as a Taylor expansion
\begin{equation}
    \mathbf{Q}(x,t+\Delta t) = \mathbf{Q}(x,t) + \Delta t \frac{\partial \mathbf{Q}}{\partial t} + \frac{1}{2} \Delta t^2 \frac{\partial^2 \mathbf{Q}}{\partial t^2} + \mathcal{O}(\Delta t^3).
\end{equation}
A method is first order in time if it only uses the term linearly dependent on $\Delta t$. The second order schemes include the terms dependent on $\Delta t^2$. A number of systems that achieve this have been developed \citep{ja:abgrall2003, ja:depalma2005, ja:rossiello2009, ja:ricchiuto2010}. This work includes extending the schemes described above to allow for second order temporal accuracy, and perform extensive studies of the various properties of the new system. Below we will summarise the extension to second order temporal accuracy, as well as the potential options that have been developed to implement this extension.

When dealing with a time dependent problem, there is clearly going to be a time dependence in the residual itself. In the first order formulation, the element residual was defined as the integral of the divergence of the flux over the element. To include the time dependence in the residual, it is necessary to define a new residual, the \textit{total residual} $\Phi^T$, which is the integral over the whole set of equations
\begin{equation}
    \begin{split}
        \Phi^T(\mathbf{Q}_h) = \int_T \left[\frac{\partial \mathbf{Q}_h}{\partial t} + \nabla \cdot \mathcal{F}_h(\mathbf{Q}_h)\right]dx dy \\
        = \int_T \frac{\partial \mathbf{Q}_h}{\partial t} dxdy + \phi^T(\mathbf{Q}_h).
    \end{split}
\end{equation}
This residual now contains a way to take into account the change in the state over the time step. There is some inconsistency in the notation and naming conventions within the residual distribution field, but here we will use the above naming scheme, where the element residual $\phi^T$ is the area integral of the divergence, and the total residual $\Phi^T$ is the integral of the whole equation. The integral over the time derivative simply becomes the mean of the time derivatives of the solution at each node of the element multiplied by the area
\begin{equation}
    \Phi^T = \sum_{j=1}^3 \frac{|T|}{3} \frac{d\mathbf{Q}_j}{dt} + \phi^T.
\end{equation}
This still contains the time derivative of the state, which is simply taken as the absolute change in the state at vertex $j$ for time step $\Delta \mathbf{Q}/\Delta t$. The distribution of this new residual requires a way to distribute the time dependent part. This is achieved by applying a mass matrix $\mathbf{m}$ \citep{ja:caraeni2002, ja:depalma2005, ja:ricchiuto2010}, which sets a fraction of the contribution from the temporal part of the total residual to be sent to each vertex. This is used to find the nodal total residual with
\begin{equation}
    \Phi^T_i = \sum_{j=1}^3 m_{ij} \frac{d \mathbf{Q}_j}{dt} + \phi^T_i.
\end{equation}
There are a number of choices for the mass matrix that offer different dissipative properties \citep{ja:ricchiuto2010}. We utilise the simplest of these, which is found by replacing the element residual in the first order method with the total residual \citep{ja:caraeni2002}. If we take the LDA scheme, the mass matrix then simply becomes
\begin{equation}
    m_{ij}^{F1} = \frac{|T|}{3}\beta_j,
\end{equation}
where $\beta_j$ is the LDA distribution matrix for the $jth$ vertex of that element. This splits the total residual in exactly the same way as the first order LDA scheme, with the addition of the temporal part to the distributed residual.


The RD method can now be recast as as the distribution of this new total residual. To achieve the second order accuracy in time, a Runge-Kutta time stepping scheme is applied. These methods function by constructing an intermediate state, and then finding the final state, for a given time step, as function of the original and intermediate states. Second (RK2), third (RK3), and fourth (RK4) order Runge-Kutta methods have been developed for the RD approach, but the additional computational costs of the RK3 and RK4 methods do not show a significant improvement in the accuracy of the numerical results \citep{ja:ricchiuto2010}. We have only considered the RK2 approach here. 


For the RD problem, for the time step $n$ to $n+1$, the intermediate state is constructed using the first order solver
\begin{equation} \label{eq:second_rd1}
    \mathbf{Q}_i^* = \mathbf{Q}_i^n - \frac{\Delta t}{|S_i|} \sum_{T|i \in T} \phi_i
\end{equation}
and the final state is found using the distribution of the total residual with
\begin{equation}
    \mathbf{Q}_i^{n+1} = \mathbf{Q}_i^* - \frac{\Delta t}{|S_i|}\sum_{T|i \in T} \Phi_i 
\end{equation}
where the total residual is calculated based on both the initial and intermediate element residuals, in the standard RK2 form \citep{ja:ricchiuto2010}. The second sub-step update becomes 
\begin{equation} \label{eq:second_rd2}
    \begin{split}
        \mathbf{Q}_i^{n+1} = \mathbf{Q}_i^* - \frac{\Delta t}{|S_i|}\sum_{T|i \in T} \left(\sum_{j=1}^3 m_{ij} \frac{\mathbf{Q}^*_i - \mathbf{Q}^n_i}{\Delta t} \right. \\ \left. + \frac{1}{2}\left(\phi_i(\mathbf{Q}^*_i) + \phi_i(\mathbf{Q}_i^n)\right)\right).
    \end{split}
\end{equation}
This provides all the information needed to construct a second order RD solver. the new total residual is only dependent on the initial state, the intermediate state, the time step, and the element residual for both the initial and intermediate states. As such there is no need to define specific second order forms for the different distribution schemes.

\subsection{Residual Distribution for the Euler Equations} \label{app:RD_derivation_Euler}
Here we describe the intricacies of transforming the Euler equations, in this case in 2D, into the appropriate form for solving by with an RD approach. This will primarily cover the linearization of the equations through the use of the Roe parameter vector \citep{ja:roe1981}.

The RD method is only applicable to linear sets of PDEs. The Euler equations are non-linear, as the Flux term is dependent on the state vector. The equations must be recast in a quasi-linear form. In order to produce the desired linearisation, a new parameter vector is defined, such that the Jacobian is only linearly dependent on the unknown of the PDE. The Roe parameter vector $\mathbf{Z}$ is suitable for this purpose, defined for the two dimensional case as 
\begin{equation}
    \mathbf{Z} = 
    \begin{pmatrix}
        \sqrt{\rho} \\
        \sqrt{\rho}v_x \\
        \sqrt{\rho}v_y \\
        \sqrt{\rho}H
    \end{pmatrix}.
\end{equation}
Starting with the original inviscid 2D Euler equations, this is converted into a PDE of $\mathbf{Z}$ by introducing the partial derivatives with respect to the new vector, $\partial \mathbf{Q}/\partial \mathbf{Z}$ and $\partial \mathbf{F}_j/\partial \mathbf{Z}$, using the chain rule. The Euler equations now appear as
\begin{equation} \label{eq:method_euler_2D_roe}
    \frac{\partial \mathbf{Q}}{\partial \mathbf{Z}}\frac{\partial \mathbf{Z}}{\partial t} + \sum_{j=1}^2 \frac{\partial \mathbf{F}_j}{\partial \mathbf{Z}}\frac{\partial \mathbf{Z}}{\partial x_j} = 0.
\end{equation}
The state vector can be rewritten in terms of the new parameter vector as
\begin{equation} \label{eq:method_euler_state_z}
    \mathbf{Q} = \left(
    \begin{matrix}
        Z_1^2 \\
        Z_1 Z_2 \\
        Z_1 Z_3 \\
        \frac{Z_1 Z_4}{\gamma} + \frac{\gamma-1}{2\gamma}(Z_2^2 + Z_3^2)
    \end{matrix}
    \right),
\end{equation}
with the flux vectors as
\begin{equation}  \label{eq:method_euler_fluxx_z}
    \mathbf{F}_x = \left(
        \begin{matrix}
            Z_1 Z_2 \\
            \frac{\gamma - 1}{\gamma} Z_1 Z_4 + \frac{\gamma + 1}{2\gamma}Z_2^2 - \frac{\gamma - 1}{2\gamma} Z_3^2 \\
            Z_2 Z_3 \\
            Z_2 Z_4
        \end{matrix}
    \right),
\end{equation}
and
\begin{equation}  \label{eq:method_euler_fluxy_z}
    \mathbf{F}_y = \left(
        \begin{matrix}
            Z_1 Z_3 \\
            Z_2 Z_3 \\
            \frac{\gamma - 1}{\gamma} Z_1 Z_4 + \frac{\gamma + 1}{2\gamma}Z_3^2 - \frac{\gamma - 1}{2\gamma} Z_2^2 \\
            Z_3 Z_4
        \end{matrix}
    \right).
\end{equation}
From these forms, it is clear that the state and flux vectors depend quadratically on the Roe vector. The partial derivatives of these vectors, with respect to the Roe parameter vector, are therefore linearly dependent on the new parameters. This means that $\mathbf{Z}$ can be used to linearise the Euler equations using the form given in equation (\ref{eq:method_euler_2D_roe}).

With the Roe parameter vector, the Jacobian in Equation (\ref{eq:method_euler_2D_roe}) satisfies the requirements given in Section \ref{sec:method_roe} for the Roe solver. The Jacobian in this case is the derivative of the flux with respect to $\mathbf{Z}$. The Jacobian at the boundary $\bar{\mathbf{A}}$, or in the element in the 2D case, is simply the Jacobian of the mean state of the vertices
\begin{equation}
    \bar{\mathbf{A}}_x = \mathbf{A}_x(\bar{\mathbf{Z}}) = \mathbf{A}_x \left(\frac{\mathbf{Z}_1 + \mathbf{Z}_2 + \mathbf{Z}_3}{3}\right).
\end{equation}
The element residual for this set of PDEs is now defined with respect to the new unknown $\mathbf{Z}$, such that Equation (\ref{eq:method_2D_res_disc}) is equivalent to
\begin{equation}
    \phi^T = 
    \frac{1}{2} \sum_{i=1}^3 \mathbf{Z}_i \frac{\partial}{\partial \mathbf{Z}} \mathcal{F}(\bar{\mathbf{Z}}_i) \cdot \mathbf{n}_i.
\end{equation}
 However, in this formulation, the residual will calculate the update to the Roe parameter vector, rather than fluid state vector $\mathbf{Q}$. In order to make use of this Roe parameter vector to update the fluid state, the fluid state must be reintroduced into the residual
\begin{equation}
    \phi^T = 
    \frac{1}{2} \sum_{i=1}^3 \mathbf{Z}_i \frac{\partial}{\partial \mathbf{Z}} \mathcal{F}(\bar{\mathbf{Z}}) \frac{\partial}{\partial \mathbf{Q}}\mathbf{Z}(\bar{\mathbf{Z}})
    \cdot 
    \mathbf{n}_i \frac{\partial}{\partial \mathbf{Z}}\mathbf{Q}(\bar{\mathbf{Z}}).
\end{equation}
These two equations are equivalent, but the second form allows us to write the residual for the original Euler equations, but calculated as a function of the mean state $\bar{\mathbf{Z}}$, rather than the fluid state. The residual is therefore given by
\begin{equation} \label{eq:method_euler_2D_res_final}
    \phi^T = \sum_{j=1}^3 \mathbf{K}_i (\bar{\mathbf{Z}}) \hat{\mathbf{Q}}_i(\bar{\mathbf{Z}}),
\end{equation}
which is directly comparable to the generic discrete form of the element residual from Equation (\ref{eq:method_2D_res_final}). The variables $\mathbf{K}_i$ and $\hat{\mathbf{Q}}_i$ of this specific discrete form are 
\begin{equation} \label{eq:method_euler_2D_qhat}
    \begin{split}
        \hat{\mathbf{Q}}_i(\bar{\mathbf{Z}}) = \frac{\partial}{\partial \mathbf{Z}}\mathbf{Q} (\bar{\mathbf{Z}})\mathbf{Z}_i =
    \left(
        \begin{matrix}
            2 \bar{Z}_1 Z_1 \\
            \bar{Z}_2 Z_1 + \bar{Z}_1 Z_2 \\
            \bar{Z}_3 Z_1 + \bar{Z}_1 Z_3 \\
            \frac{1}{\gamma}\left(\bar{Z}_4 Z_1 + \gamma_1 \bar{Z}_2 Z_2 + \gamma_1 \bar{Z}_3 Z_3 + \bar{Z}_1 Z_4 \right)
        \end{matrix}
    \right),
    \end{split}
\end{equation}
and
\begin{equation}
    \begin{split}
         \mathbf{K}_i(\bar{\mathbf{Z}}) = \frac{1}{2}\frac{\partial }{\partial \mathbf{Z}}\mathcal{F}(\bar{\mathbf{Z}})\frac{\partial}{\partial \mathbf{Q}}\mathbf{Z}(\bar{\mathbf{Z}}) \cdot \mathbf{n}_i \\ = \frac{1}{2} \frac{\partial }{\partial \mathbf{Q}} \mathcal{F}(\bar{\mathbf{Z}}) \cdot \mathbf{n}_i \\ = \frac{1}{2} \mathcal{A}(\bar{\mathbf{Z}})\cdot \mathbf{n}_i.
    \end{split}
\end{equation}
This $\mathbf{K}_i$ matrix is sometimes referred to as the inflow matrix, as it can be used to encode the nature of the flow at each vertex since it projects the Jacobian onto the normal of the face opposite that vertex. The $\mathbf{K}_i$ matrix is now defined as the average of Jacobian matrices of the original form of the Euler equations, projected onto the edge normals of the element, where
\begin{equation}
    \mathcal{A}(\bar{\mathbf{Z}}) = \left(\bar{\mathbf{A}}_x(\bar{\mathbf{Z}}), \bar{\mathbf{A}}_y(\bar{\mathbf{Z}}) \right) = \left(
    \frac{\partial}{\partial \mathbf{Q}} \mathbf{F}_x(\bar{\mathbf{Z}}),
    \frac{\partial}{\partial \mathbf{Q}} \mathbf{F}_y(\bar{\mathbf{Z}})\right).
\end{equation}
It is important to note that the Jacobian is being calculated for the average Roe parameter, which produces a subtly different result to using the average fluid state. The introduction of $\hat{\mathbf{Q}}_i$, which has the same units as the fluid state but clearly differs from it in detail, and the fact that the Jacobian is evaluated at the average Roe state, together encode the effect of the linearisation.

To summarise, the RD method is only applicable to linear sets of PDEs, so a suitable linearisation of the Euler equations is required. Roe produced such a linearisation, initially for the Roe solver, but it is usable in this context as well. To calculate the residual for use in the update of the fluid state, the $K$ matrix is based on the Jacobians evaluated at the average Roe parameter for the spatial element. This is combined with a variable that is analogous to the state variable, but defined with the chosen linearisation. These produce a consistent definition of the residual for that can be used to update the fluid state, without losing the effects of the linearisation and invalidating the scheme. The exact form of the decomposed $\bm{K}_i$ is given in the appendices of \citet{ja:paardekooper2017}.

\section{3D $\mathbf{K}$-matrix} \label{app:3D_kmatrix}
The 3D K-matrix, at vertex $i$, is defined similarly to the 2D case \citep{ja:paardekooper2017}, where $K_i=(A_x n_x + A_y n_y + A_z n_z)/3$, where $\bm{n}_i=(n_x,n_y,n_z)$ is the unit normal of the face opposite vertex $i$, and $A = (A_x,A_y,A_z)$ are the triangle Jacobians in each dimension. However, the form found by finding the arithmetic mean of the Jacobian projections onto the respective normals does not allow for the construction of the corresponding $K^-$ and $K^+$ matrices. These require selections by negative/positive eigenvalues respectively. The desired form is found by decomposing the initial form of the K-matrix into its Schur decomposition form, with $\mathbf{K}_i = \mathbf{R}^{-1} \mathbf{\Lambda} \mathbf{R}$, where $\mathbf{\Lambda}$ is a diagonal matrix, with the eigenvalues as the diagonal elements. This is the matrix product of $\mathbf{\Lambda}$, which is the diagonal matrix composed of the eigenvalues of $\mathbf{K}_i$, and the right hand matrix $\mathbf{R}$, made up of columns consisting of the eigenvectors of the K-matrix. The end result is a form that can be used to calculate the inflow matrix $\bm{K}_i$, or its negative or positive counterparts, given as
\begin{multline*}
    K_{11} = \frac{\alpha_c}{c}\lambda_{123} - \frac{\Omega}{c} \lambda_{12} + \lambda_3, \\
    K_{12} = -\frac{\gamma_1 v_{xc}}{c}\lambda_{123} + \frac{n_x}{c}\lambda_{12}, \\
    K_{13} = -\frac{\gamma_1 v_{yc}}{c}\lambda_{123} + \frac{n_y}{c}\lambda_{12}, \\
    K_{14} = -\frac{\gamma_1 v_{zc}}{c}\lambda_{123} + \frac{n_z}{c}\lambda_{12}, \\ 
    K_{15} = \frac{\gamma_1}{c^2}\lambda_{123}, \\
\end{multline*}
\begin{multline*}
    K_{21} = (\alpha_c v_{xc} - \Omega n_x)\lambda_{123} + (\alpha_c n_x - v_{xc}\Omega)\lambda_{12}, \\
    K_{22} = (n_x^2-\gamma_1 v_{xc}^2)\lambda_{123} - \gamma_2 v_{xc} n_x \lambda_{12} + \lambda_3, \\
    K_{23} = (n_x n_y - \gamma_1 v_{xc} v_{yc})\lambda_{123} + (v_{xc} n_y - \gamma_1 v_{yc} n_x)\lambda_{12}, \\
    K_{24} = (n_x n_z - \gamma_1 v_{xc} v_{zc})\lambda_{123} + (v_{xc} n_z - \gamma_1 v_{zc} n_x)\lambda_{12}, \\
    K_{25} = \frac{\gamma_1 v_{xc}}{c}\lambda_{123} + \frac{\gamma_1 n_x}{c}\lambda_{12}, \\
\end{multline*}
\begin{multline*}
    K_{31} = (\alpha_c v_{yc} - \Omega n_y)\lambda_{123} + (\alpha_c n_y - v_{yc}\Omega)\lambda_{12}, \\
    K_{32} = (n_x n_y - \gamma_1 v_{xc} v_{yc})\lambda_{123} + (v_{yc} n_x - \gamma_1 v_{xc} n_y)\lambda_{12}, \\
    K_{33} = (n_y^2-\gamma_1 v_{yc}^2)\lambda_{123} - \gamma_2 v_{yc} n_y \lambda_{12} + \lambda_3, \\
    K_{34} = (n_y n_z - \gamma_1 v_{yc} v_{zc})\lambda_{123} + (v_{yc} n_z - \gamma_1 v_{zc} n_y)\lambda_{12}, \\
    K_{35} = \frac{\gamma_1 v_{yc}}{c}\lambda_{123} + \frac{\gamma_1 n_y}{c}\lambda_{12}, \\
\end{multline*}
\begin{multline*}
    K_{41} = (\alpha_c v_{zc} - \Omega n_z)\lambda_{123} + (\alpha_c n_z - v_{zc}\Omega)\lambda_{12}, \\
    K_{42} = (n_x n_z - \gamma_1 v_{xc} v_{zc})\lambda_{123} + (v_{zc} n_x - \gamma_1 v_{xc} n_z)\lambda_{12}, \\
    K_{43} = (n_y n_z - \gamma_1 v_{yc} v_{zc})\lambda_{123} + (v_{zc} n_y - \gamma_1 v_{yc} n_z)\lambda_{12}, \\
    K_{44} = (n_z^2-\gamma_1 v_{zc}^2)\lambda_{123} - \gamma_2 v_{zc} n_z \lambda_{12} + \lambda_3, \\
    K_{45} = \frac{\gamma_1 v_{zc}}{c}\lambda_{123} + \frac{\gamma_1 n_z}{c}\lambda_{12}, \\
\end{multline*}
\begin{multline*}
    K_{51} = (\alpha_c H_c - \Omega^2) \lambda_{123} + \Omega(\alpha_c - H_c)\lambda_{12}, \\
    K_{52} = (\Omega n_x - v_x - \alpha_c v_{xc})\lambda_{123} + (H_c n_x - \gamma_1 v_{xc} \Omega)\lambda_{12}, \\
    K_{53} = (\Omega n_y - v_y - \alpha_c v_{yc})\lambda_{123} + (H_c n_y - \gamma_1 v_{yc} \Omega)\lambda_{12}, \\
    K_{54} = (\Omega n_z - v_z - \alpha_c v_{zc})\lambda_{123} + (H_c n_z - \gamma_1 v_{zc} \Omega)\lambda_{12}, \\
    K_{55} = \frac{\gamma_1 H_c}{c}\lambda_{123} + \frac{\gamma_1 \Omega}{c}\lambda_{12} + \lambda_3, \\
\end{multline*}
where $X_c \equiv X/c$. The new $\lambda$ terms have the same meaning as in the 2D case, with $\lambda_{123} = (\lambda_1 + \lambda_2 - 2\lambda_3)/2$ and $\lambda_{12} = (\lambda_1-\lambda_2)/2$. Similarly to the 2D form, the eigenvalues used here are $\lambda_1=\Omega + c$, $\lambda_1=\Omega - c$, and $\lambda_3=\lambda_4=\lambda_5=\Omega$. The new variables are $\alpha = \gamma_1 (v_x^2+v_y^2+v_z^2)/2$ and $\Omega = v_x n_x + v_y n_y + v_z n_z$. To find the corresponding $\mathbf{K}_i^+$ and $\mathbf{K}_i^-$, one simply uses only the positive $\lambda^+$ or negative $\lambda^-$ eigenvalues, where
\begin{equation}
    \lambda_i^+ = \max(0,\lambda_i) \hspace{5mm} \mathrm{and} \hspace{5mm} \lambda_i^- = \min(0,\lambda_i).
\end{equation}
Together this matrix form, and the associated eigenvalues, entirely describe the inflow matrix, allowing us to implement the 3D form of the RD hydro solver.


\bsp	
\label{lastpage}
\end{document}